\documentclass[%
 amsmath,amssymb,
prl,%
twocolumn,
superscriptaddress
]{revtex4-1}

\usepackage{graphicx}% Include figure files
\usepackage{tabularx}
\usepackage{dcolumn}% Align table columns on decimal point
\usepackage{bm}% bold math
\usepackage{hyperref}
\usepackage[utf8]{inputenc}
\usepackage[T1]{fontenc}
\usepackage{mathptmx}
\usepackage{url}
\usepackage{xcolor}

\begin{document}

%\preprint{AIP/123-QED}

\title{Promising photovoltaic efficiency of a layered silicon oxide crystal Si\texorpdfstring{\textsubscript{3}}{3}O}

\author{Sejoong Kim}
 \affiliation{University of Science and Technology (UST), Daejeon 34113, Korea}
 \affiliation{Korea Institute for Advanced Study, Seoul 02455, Korea}

\author{Kisung Chae}%
\email{corresponding author: kisung@kias.re.kr}
\affiliation{Korea Institute for Advanced Study, Seoul 02455, Korea}
\affiliation{Department of Chemistry and Biochemistry, University of California, San Diego,La Jolla, CA 92093, United States}
\affiliation{Materials Science and Engineering Department, The University of Texas at Dallas, Richardson, Texas 75080, United States}

\author{Young-Woo Son}
 \email{corresponding author: hand@kias.re.kr}
 \affiliation{Korea Institute for Advanced Study, Seoul 02455, Korea}

\date{\today}

\begin{abstract}
Computational searching and screening of new functional materials exploiting earth abundant elements
can accelerate developments of their energy applications.
Based on a state-of-the-art materials search algorithm and {\it ab initio} calculations, 
we demonstrate a recently suggested stable silicon oxide 
with a layered structure (Si$_{3}$O) as an ideal photovoltaic material. With many-body first-principles approaches, 
the monolayer and layered bulk of Si$_{3}$O show direct quasiparticle gaps of 1.85 eV and 1.25 eV, respectively, 
while an optical gap of about 1.2 eV is nearly independent of the number of layers. 
Spectroscopic limited maximum efficiency (SLME) is estimated to be 27\% for a thickness of 0.5 $\mu$m, 
making it a promising candidate for solar energy applications.
\end{abstract}

\maketitle

Discovery of efficient, reliable and safe photovoltaic (PV) materials
will be a critical booster to realize a sustainable energy alternative with solar cells~\cite{Haegel141}.
%will be a critical booster to realize solar energy harvesting applications, serving as a sustainable energy source. % KC
Among various potential candidates for the next generation solar cells,
two-dimensional (2D) materials have gained significant attention for their few-atom thickness, 
exceptional stability and diverse electronic structures 
tunable by the number of layers and heterostructure 
formations~\cite{StranoNatureNano,C4NR01600A, Bonaccorso1246501,Novoselovaac9439,acsphotonics.7b01103,adma.201802722}. 
To uncover their full potentials, it is important to search and test 2D materials 
candidates thoroughly~\cite{PhysRevLett.118.106101,Marzari2018}. 

In discovering materials {\it in silico}, various computational materials design methods have been adopted to accelerate the discovery. 
Crystal structure predictions based on global optimization~\cite{JM9950501269,Oganov,PhysRevB.82.094116}
have been used to predict novel materials 
with unusual properties~\cite{kim_synthesis_2015,davies_computer-aided_2018,gu_prediction_2017}.
Together with these impressive developments in materials predictions, 
recent progresses in the data-driven sciences 
such as text-mining and natural language processing for synthesis of inorganic materials~\cite{kononova_text-mined_2019}
may further accelerate the computer-aided materials discovery.
For 2D materials, 
high-throughput materials search algorithms~\cite{PhysRevLett.118.106101,Marzari2018,PhysRevLett_108_068701_2012, acs.chemmater.9b02166} have been used for screening the optimal material candidates for solar cell applications~\cite{PhysRevLett_108_068701_2012, acs.chemmater.9b02166, wang_kagse_2018, yang_monolayer_2019,zhao_design_2017}.
Considering current domination of silicon solar cells, however,
a suitable 2D silicon material that is readily applicable for 
harvesting solar light is still lacking. 

\begin{figure}[t]
\begin{center}
\includegraphics[width=0.95\columnwidth, clip=true]{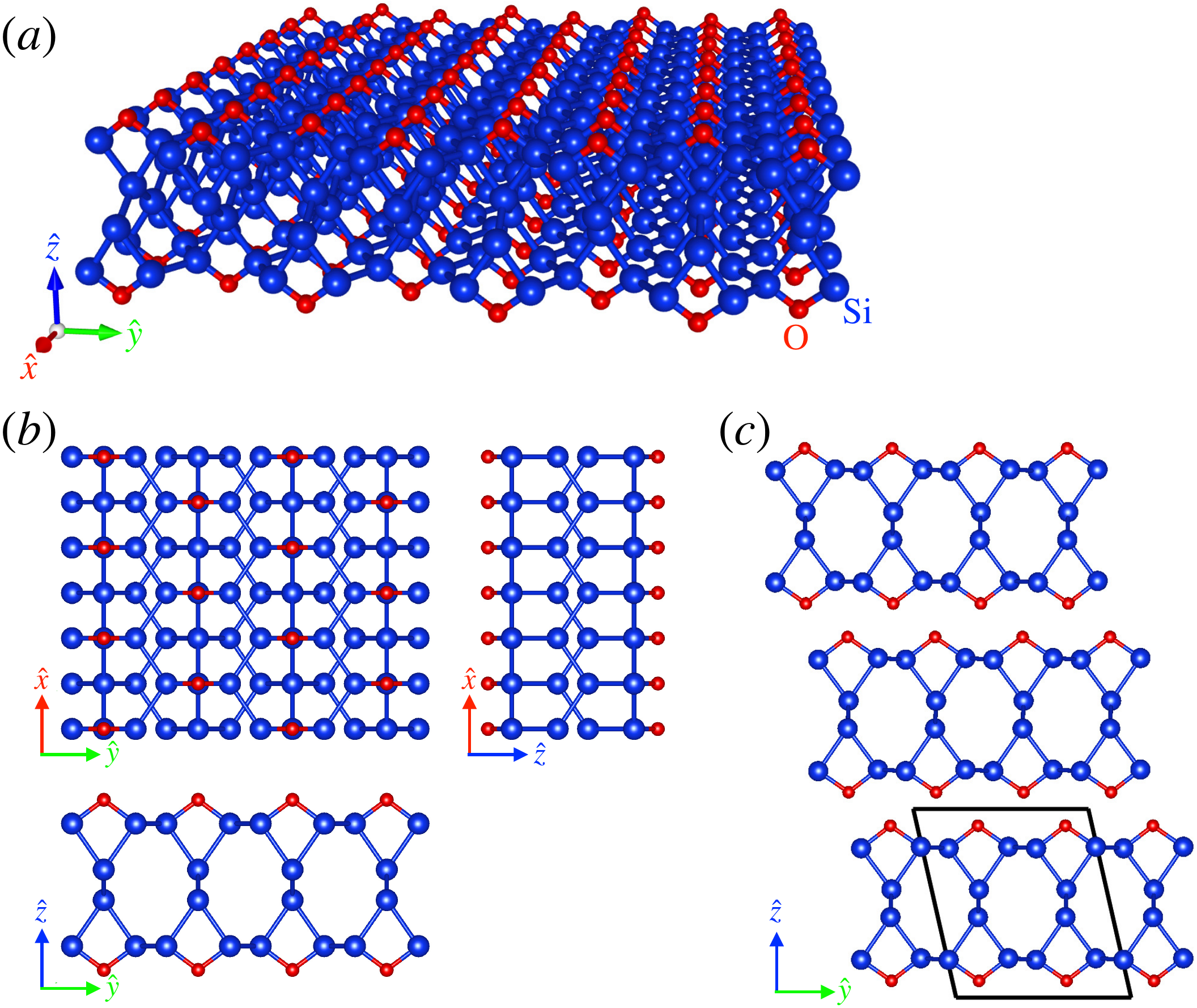} 
\end{center}
\caption{\label{config} (a) Schematic atomic configuration of 
monolayer structure of $\textrm{Si}_{3}\textrm{O}$. Red and blue balls denote
oxygen and silicon atoms, respectively.
(b) Top and side views of the monolayer. 
(c) The stacking configuration of bulk $\textrm{Si}_{3}\textrm{O}$. 
Black solid lines indicate the unit cell of the layered bulk. 
Atomic configurations are generated by $\textsc{Vesta}$~\cite{IUCr_Newslett_7106-119_2006}.}
\end{figure}

Recently, a family of novel 2D materials composed of group IV and VI elements, namely T$_{3}$X (T=C, Si, Ge, Sn; X=O, S, Se, Te), have been predicted, demonstrating various electronic properties ranging from band insulator to quantum spin Hall insulator~\cite{NanoLett_19_2694_2019}. Those novel 2D crystals have been found by using a new crystal structure prediction method: \textsc{Sandwich} (Search by \emph{Ab initio} 
Novel Design via Wyckoff positions Iteration in Conformational Hypersurface)~\cite{NanoLett_19_2694_2019,chae_new_2018}. 
This method has a particular merit over the others in finding stable but unconventional 2D atomic structures with a low atomic density, 
e.g., hollow structures~\cite{NanoLett_19_2694_2019,chae_new_2018}.
Among the T$_{3}$X compounds, %the 
2D Si$_{3}$O crystal shown in Fig.~\ref{config} has the convex-hull stability and a direct band gap, making it a good candidate for optoelectronic devices~\cite{NanoLett_19_2694_2019,chae_new_2018}.

In this work, 
we provide computational evidences showing that Si$_3$O has noticeable merits for solar cell application.
Our evaluation of the spectroscopic limited maximum efficiency (SLME)~\cite{PhysRevLett_108_068701_2012}
of Si$_3$O reaches 26.8\% for a thickness of 0.5 $\mu$m. 
As a good metric for screening PV materials~\cite{PhysRevLett_108_068701_2012}, % KC
the SLME was also used as a key descriptor to test over a million materials in a machine-learning approach, searching for solar energy application~\cite{acs.chemmater.9b02166}. % KC
The SLME value in this study is comparable to the best 2D candidates in the database of Ref.~\cite{acs.chemmater.9b02166}.
To obtain an accurate value, we perform fully converged many-body 
{\it ab initio} calculations on quasiparticle (QP) spectra 
and exciton binding energies of a monolayer and bulk, respectively. 
It is shown that the both have direct gaps, differing 
from other semiconducting 2D materials~\cite{PhysRevLett.105.136805}. 
Considering other merits of Si$_3$O such as earth-abundant, non-toxic and simple elements, light weight,
a superb stability~\cite{NanoLett_19_2694_2019} and a very light effective mass~\cite{chae_new_2018}, 
we expect that the proposed layered silicon oxide 
will play an important role in silicon-based energy applications
once it be synthesized. 
We briefly discuss synthesis and stability of the Si$_{3}$O in the Electronic Supplementary Information (ESI).

%%%%%%%%%%%%%%%%%%%%%%%%%%%%%%%%%%%%%%%%%%%%%%%%%%%%%%%%%%%%%%%%%%%%%%%%%%%%%
% Figure: Si3O structure
% (a) monolayer
% (b) bulk
%%%%%%%%%%%%%%%%%%%%%%%%%%%%%%%%%%%%%%%%%%%%%%%%%%%%%%%%%%%%%%%%%%%%%%%%%%%%%
%%%%%%%%%%%%%%%%%%%%%%%%%%%%%%%%%%%%%%%%%%%%%%%%%%%%%%%%%%%%%%%%%%%%%%%%%%%%%

We use atomic structures of a monolayer 
and layered bulk of Si$_3$O obtained 
from our \textsc{Sandwich} code, % KC 
as shown in Figs.~\ref{config}(b) and \ref{config}(c), respectively~\cite{NanoLett_19_2694_2019,chae_new_2018}.
We perform the mean-field~\cite{PhysRevLett_77_3865_1996} density functional theory (DFT-PBE) calculations 
using \textsc{Quantum Espresso}~\cite{JPhys_CM_21_395502_2009,JPhys_CM_29_465901_2017}. 
Grimme's DFT-D2 method~\cite{J_Comp_Chem_27_1787_2006} 
is used for interlayer interactions. % KC
On top of the DFT-PBE calculations, the $G_0W_0$~\cite{PhysRev_139_A796_1965, PhysRevB_87_165124_2013} and Bethe-Salpeter equation (BSE)~\cite{PhysRev_84_1232_1951, PhysRevLett_80_4510_1998, PhysRevLett_80_4514_1998, PhysRevB_34_5390_1986,PhysRevLett_81_2312_1998,PhysRevB_62_4927_2000}
calculations are performed by using \textsc{BerkeleyGW}~\cite{Com_Phys_Comm_183_1269_2012}. 
The Coulomb interaction truncation~\cite{PhysRevB_73_233103_2006} is used to simulate the isolated monolayer. Due to a slow convergence of the QP energies in 2D systems~\cite{PhysRevLett_111_216805_2013}, % KC
we carefully converge the QP energy gap within $50$ meV 
by varying vacuum size, $k$-point grid, energy cutoff for dielectric matrix and the number of unoccupied bands. % KC
We also calculate self-consistent Green's function to obtain QP gaps, i.e., $GW_0$ approximation. % KC
It is known that the band gaps obtained by the $GW_0$ method converge to the QP gap linearly interpolated from the $G_0W_0$ gap~\cite{PhysRevB_34_5390_1986}. 
After three iterations, the difference between them becomes smaller than $5$ meV. 
Further details of calculations can be found in the ESI.

%\section{Results and Discussion}
%\subsection{QP Band Structure}
%%%%%%%%%%%%%%%%%%%%%%%%%%%%%%%%%%%%%%%%%%%%%%%%%%%%%%%%%%%%%%%%%%%%%%%%%%%%%
% Figure: DFT vs GW bands 
% (a) monolayer
% (b) bulk
%%%%%%%%%%%%%%%%%%%%%%%%%%%%%%%%%%%%%%%%%%%%%%%%%%%%%%%%%%%%%%%%%%%%%%%%%%%%%
\begin{figure}[ht!]
\begin{center}
\includegraphics[width=0.95\columnwidth, clip=true]{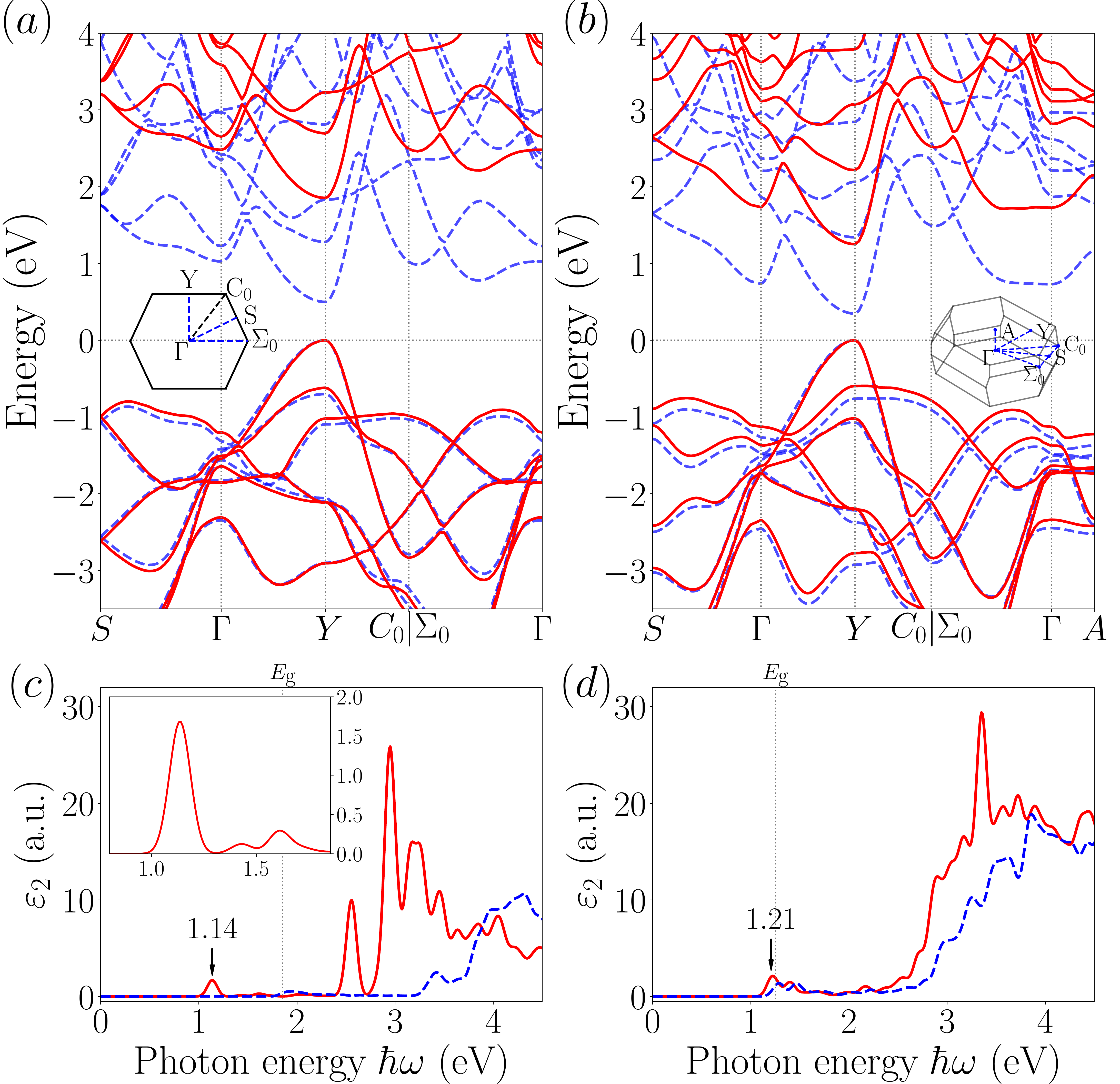} 
\end{center}
\caption{\label{bands} Energy band structures of (a) monolayer and (b) bulk of $\textrm{Si}_{3}\textrm{O}$. 
Blue dashed and red solid lines represent energy bands calculated from  $\textrm{DFT-PBE}$ and $GW$, respectively. 
The valence band maximum is set to be zero. Insets show symmetric points in the 1st Brillouin zone.
Absorption spectra $\epsilon_{2}$ of (c) monolayer and (d) bulk from $GW\textrm{+BSE}$ (red solid line) and $GW\textrm{+RPA}$ (blue dashed line). 
The monolayer spectrum $\epsilon_{2}$ is normalized by the thickness of the Si$_{3}$O relative to the corresponding cell including vacuum padding. The effective thickness of the monolayer is obtained by estimating the distance of two planes parallel to the monolayer Si$_{3}$O, between which 99\% of the charge density in included.
Dotted gray lines indicate photon energy equal to the QP energy gap. 
The brightest excitons are indicated by black arrows with their energies. 
The inset in (c) is an enlarged view near 1 eV.}
\end{figure}
%%%%%%%%%%%%%%%%%%%%%%%%%%%%%%%%%%%%%%%%%%%%%%%%%%%%%%%%%%%%%%%%%%%%%%%%%%%%%

%%%%%%%%%%%%%%%%%%%%%%%%%%%%%%%%%%%%%%%%%%%%%%%%%%%%%%%%%%%%%%%%%%%%%%%%%%%%%
% Table: Band gap calculations
% DFT+PBE, HSE06, G0W0, GW0
%%%%%%%%%%%%%%%%%%%%%%%%%%%%%%%%%%%%%%%%%%%%%%%%%%%%%%%%%%%%%%%%%%%%%%%%%%%%%
\begin{table}[b]
\small
  \caption{\ Direct band gap energies $E_{g}$ at $\textrm{Y}$ for monolayer and bulk $\textrm{Si}_{3}\textrm{O}$ calculated by DFT+PBE, hybrid functional HSE06~\cite{JChemPhys_118_8207_2003,JChemPhys_125_224106_2006}, 
$G_{0}W_{0}$ and $GW_{0}$ approximations.}
  \label{band_gaps}
  \begin{tabular*}{0.48\textwidth}{@{\extracolsep{\fill}}lll}
    \hline
    calculation method & monolayer & bulk \\
    \hline
    DFT+PBE      & 0.50 eV & 0.35 eV \\
    HSE06~\cite{NanoLett_19_2694_2019}        & 1.20 eV & 1.03 eV \\
    $G_{0}W_{0}$ & 2.18 eV & 1.47 eV \\
    $GW_{0}$     & 1.85 eV & 1.25 eV \\
    \hline
  \end{tabular*}
\end{table}
%%%%%%%%%%%%%%%%%%%%%%%%%%%%%%%%%%%%%%%%%%%%%%%%%%%%%%%%%%%%%%%%%%%%%%%%%%%%%

Figures~\ref{bands}(a) and \ref{bands}(b) show band structures of monolayer and bulk $\textrm{Si}_{3}\textrm{O}$, respectively, by using both DFT-PBE and $GW_0$ calculations.
Both systems have direct band gaps ($E_{g}$) at the symmetric point of $\textrm{Y}$.
Gaps from other computational methods are summarized in the Table~\ref{band_gaps}. 
Note that the bandgap energies are improved to some extent by using HSE06 functional when compared to the PBE-DFT case, but they are still underestimated by
significant amount compared to the quasiparticle bandgaps, indicating self-energy corrections are necessary to investigate optical properties of layered materials.
As discussed above, the converged many-body QP gaps can be obtained 
at the level of the $GW_{0}$ approximation. 
The difference in band gaps ($\Delta E_{g}$) between DFT-PBE and $GW_0$ (or self-energy corrections) are $1.35$ eV and $0.90$ eV for monolayer and bulk, respectively. 
The different $\Delta E_{g}$ originates from the weaker screening of Coulomb interaction in 
the monolayer $\textrm{Si}_{3}\textrm{O}$ than in the bulk.
We expect efficient light absorption of Si$_{3}$O regardless of the number of layers because both monolayer and bulk Si$_{3}$O are direct band gap semiconductors. 
This is in sharp contrast to MoS$_{2}$ which undergoes a direct to indirect band gap transition and a strong photoluminescence of monolayer vanishes when it becomes bilayer or thicker~\cite{PhysRevLett.105.136805}. 
The robust direct optical absorption regardless of thickness demonstrated in Si$_3$O here will be very beneficial in PV applications.

%\subsection{Optical Properties}
We solve the BSE with QP energy bands to investigate optical properties of $\textrm{Si}_{3}\textrm{O}$ such as frequency-dependent optical absorption spectrum $\epsilon_{2}(\omega)$ (the imaginary part of the dielectric function) and exciton energy levels. Solving the BSE depends on two calculation parameters: the $k$-point density and the number of valence and conduction bands for optical transition.
We found that $40\times40\times1$ (monolayer) and $20\times20\times4$ (bulk) $k$-point grids and six bands from the valence and conduction band edges suffice to converge $\epsilon_{2}(\omega)$ up to $4.0$ eV. % KC
Figures~\ref{bands}(c) and \ref{bands}(d) show absorption spectra $\varepsilon_{2}(\omega)$ 
for monolayer and bulk, respectively. 
The absorption spectra obtained with the electron-hole interaction ($GW\textrm{+BSE}$) 
are compared with the non-interacting case ($GW\textrm{+RPA}$). 
We note that the $GW\textrm{+RPA}$ spectra for both monolayer and bulk
do not show prominent optical absorption in energy below $E_{g}$, 
marked as the dashed line in Figs.~\ref{bands}(c) and \ref{bands}(d). 
Rather, a significant amount of optical absorption appears when the photon energy exceeds at least about $1.5$ eV above $E_{g}$.
This is due to the energy landscape of the charge carriers over the whole Brillouin zone. 
As shown in Figs.~\ref{bands}(a) and \ref{bands}(b), the direct band gap $E_{g}$ occurs at the $\textrm{Y}$ point, 
while direct electron-hole pair generation at other momentum would require about 1.5--2.0 eV higher energy than the gap. 

When the electron-hole interaction is taken into account, the optical absorption is significantly changed, contrast to the non-interacting case. First, the major absorption peaks are shifted down to the lower photon energy range. 
This overall red shift of the $GW\textrm{+BSE}$ spectrum with respect to the $GW\textrm{+RPA}$ is more evident in the monolayer than the bulk as shown in Figs.~\ref{bands}(c) and \ref{bands}(d). 
As discussed above, a 2D system will have less Coulomb screening along the out-of-plane direction, so its optical absorption can be more strongly affected. Similar to MoS$_2$~\cite{PhysRevB.93.235435}, the reduced screening effects in a monolayer limit affect band structures and optical spectra in the opposite way.
Monolayer has the larger electronic band gap than bulk (1.85 eV vs 1.25 eV, respectively), but the difference is compensated by the stronger red shift in the absorption spectra (0.8 eV vs 0.2 eV), resulting in almost constant optical band gap of about 1.2 eV regardless of the thickness.

We obtain electron-hole bound states and their energy levels from the BSE. 
Monolayer shows several optically active exciton states, or bright excitons, below the optical transition continuum in the absorption spectra as shown in Fig.~\ref{bands}(c).
The main absorption peak located at $1.14$ eV corresponds to the lowest exciton level around $\textrm{Y}$ point with a significant binding energy of $0.72$ eV. 
Other exciton peaks with relatively smaller contributions within the band gap, are depicted in the inset of Fig.~\ref{bands}(c). In contrast, the bulk $\textrm{Si}_{3}\textrm{O}$ shows a single bright exciton peak at $1.21$ eV inside the band gap. 
The bright exciton at $\textrm{Y}$ is located right below the inter-band transition continuum, implying small binding energy of $40$ meV due to the increased screening effect in bulk. 

%%%%%%%%%%%%%%%%%%%%%%%%%%%%%%%%%%%%%%%%%%%%%%%%%%%%%%%%%%%%%%%%%%%%%%%%%%%%%
% Figure: Absorption vs Air Mass 1.5
% (a) Absorption (monolayer)
% (b) Absorption (bulk)
%%%%%%%%%%%%%%%%%%%%%%%%%%%%%%%%%%%%%%%%%%%%%%%%%%%%%%%%%%%%%%%%%%%%%%%%%%%%%
\begin{figure}[ht]
\begin{center}
\includegraphics[width=0.95\columnwidth, clip=true]{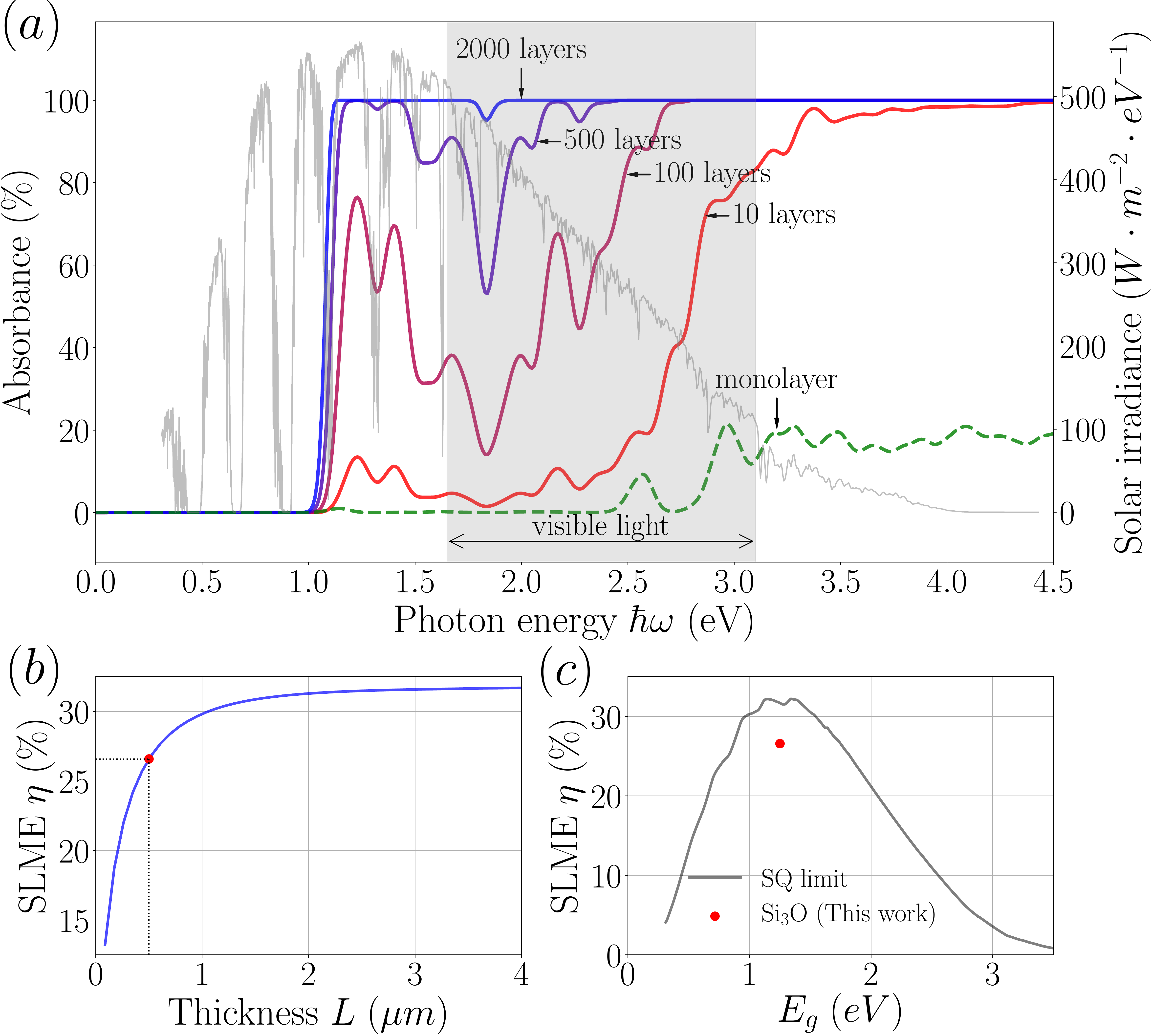} 
\end{center}
\caption{\label{absorbance} (a) Absorbance spectra from $GW\textrm{+BSE}$ for monolayer and bulk $\textrm{Si}_{3}\textrm{O}$ of 10, 100, 500, and 2000 layers whose thicknesses in $\mu m$ are $8.208\times10^{-4}$, $8.695\times10^{-3}$, $8.695\times10^{-2}$, $0.435$, and $1.739$, respectively. The gray box indicates the energy range of visible light. The AM1.5 is shown with gray lines. (b) The SLME ($\eta$) as a function of thickness $L$ at $25^{\circ}\textrm{C}$. (c) $\eta$ at $L=0.5 \;\mu m$ is shown together with the SQ limit curve as a function of the band gap.
}
\end{figure}
%%%%%%%%%%%%%%%%%%%%%%%%%%%%%%%%%%%%%%%%%%%%%%%%%%%%%%%%%%%%%%%%%%%%%%%%%%%%%

%\subsection{Photovoltaic Efficiency}
PV efficiency of $\textrm{Si}_{3}\textrm{O}$ is estimated by calculating the SLME~\cite{PhysRevLett_108_068701_2012} based on the absorption property.
The Shockley-Queisser (SQ) limit~\cite{JApplPhys_32_510_1961} is an ideal case that photons with energy greater than $E_{g}$ are perfectly absorbed.
However, in reality, light absorption varies with energy according to the absorption coefficient $\alpha(\omega)$.
The $\alpha(\omega)$ can be calculated by using the $\varepsilon_{2}(\omega)$ obtained from $GW$+BSE as follows:
\begin{equation}
\alpha(\omega) = \frac{\omega}{c\tilde{n}(\omega)}\varepsilon_{2}(\omega),
\end{equation}
where $\tilde{n}(\omega)$ is the real part of the refractive index, $\tilde{n}(\omega) = {\scriptscriptstyle \sqrt{\left(\varepsilon_{1}(\omega) + \sqrt{\varepsilon_{1}^{2}(\omega) + \varepsilon_{2}^{2}(\omega)}\right)/2}}$.
Here, real and imaginary parts of the dielectric function, $\varepsilon_{1}(\omega)$ and $\varepsilon_{2}(\omega)$, respectively, are related to each other by the Kramers-Kr\"{o}nig relation~\cite{Condensed_Matter_Physics_Marder}.
The light absorption is included in the SLME as the absorbance $A(\omega)$, which is given by 
\begin{equation}
A(\omega) = 1 - e^{-2\alpha(\omega)L}
\label{absorb}
\end{equation}
where $L$ is the material thickness. 
$A(\omega)$ is derived by assuming both no reflection of a normal incident light at the front surface and perfect reflection at the back surface of the photovoltaic material~\cite{PhysRevLett_108_068701_2012}. 
Due to the unity reflection~\cite{PhysRevLett_108_068701_2012}, the net distance traveled by light within the material is $2L$.

Figure~\ref{absorbance}(a) shows calculated absorbance spectra for monolayer and bulk $\textrm{Si}_{3}\textrm{O}$ 
with different thicknesses. 
The absorbance $A(\omega)$ in Eq.~(\ref{absorb}) increases with the $L$
because there is %more 
an increasing
chance to be absorbed 
when light travels further inside the material. 
We note, however, that the monolayer Si$_{3}$O can demonstrate desirable optoelectronic properties for particular applciations; 
it shows a great sunlight absorption ranging 10--22 \% in the visible range, which is comparable to monolayer MoS$_{2}$ (about 10--20 \% with unity reflection at the back surface)~\cite{bernardi_extraordinary_2013}, and extremely light effective masses for both electrons and holes ($\sim$0.03 $m_{0}$ with $m_{0}$ being an electron rest mass) as shown in Fig.~S3 in the ESI, indicating its good transport properties.
All in all, the monolayer Si$_{3}$O is promising for a semitransparent solar cell or flexible transparent conducting electrode. 
Furthermore, contrast to other 2D materials (e.g., MoS$_{2}$), the electronic structures show direct optical transitions regardless of the number of layers, which enables the Si$_{3}$O to be directly used as a macroscopic active layer in PV devices.

When the film is very thin (10 layers), the absorbance spectrum [Fig.~\ref{absorbance}(a)] looks similar to the absorption spectrum shown in Fig.~\ref{bands}(d). For 100 layers, the absorbance in the high energy regime ($> 2.7$ eV) becomes saturated to be unity. 
In contrast to the 10 layer case, drastically enhanced absorbance is shown even in the low energy range from $E_{g}$ to about $2.5$ eV due to the increased thickness. % KC
In particular, two dominant peaks below 1.5 eV reaches about 60 \%. % KC
As thickness further increases, the absorbance approaches to unity in the wider range of energy. 
The absorbance in the vicinity of $1.8$ eV, which corresponds to the absorption coefficient minimum, increases relatively slowly, but it finally approaches to unity when bulk $\textrm{Si}_{3}\textrm{O}$ is thick enough. 

We also calculate the SLME as a function of thickness to investigate how the absorbance affects the PV efficiency. 
Given the total incident solar energy density $P_{\textrm{in}}$ based on the Air Mass 1.5 data~\cite{AM1.5}, 
the SLME of $\eta=\textrm{max}\left[P\right]/P_{\textrm{in}}$ is obtained by 
maximizing the output power density $P=JV$, the product of the net current density $J$ 
and voltage $V$~\cite{PhysRevLett_108_068701_2012}.
Figure~\ref{absorbance}(b) shows that the SLME reaches $30.4 \%$ at thickness of about $1.8\;\mu m$, comparable to the SQ limit of $31.67\%$.
In case of $L=0.5\;\mu m$ and the temperature $T=25^{\circ}\textrm{C}$, the standard thickness for comparisons~\cite{PhysRevLett_108_068701_2012, acs.chemmater.9b02166},
the SLME is about $26.81\%$, 
which is shown on top of the SQ limit in Fig.~\ref{absorbance}(c). 

%\section{Conclusions}
In conclusion, we have shown that the layered Si$_3$O has a favorable optical gap for solar light absorption that is nearly independent of the number of layers. Its computed PV efficiency is also comparable to the ideal limit.
Thus, we believe that the new two-dimensional silicon oxide 
suggested here possesses highly desirable characteristics for solar cell, composed of a single element and oxygen which is quite analogous to the currently dominant solar cell material. % KC

\section*{Conflicts of interest}
There are no conflicts of interest to declare.

\section*{Acknowledgement}
S. K. was supported by the KISTI National Supercomputing Center with supercomputing resources including technical support (KSC-2019-CRE-0040) and by the Open KIAS Center at Korea Institute for Advanced Study. 
Y.-W.S. was supported by the NRF of Korea (Grant No. 2017R1A5A1014862, SRC program: vdWMRC Center)
and by KIAS individual grant (CG031509).

\clearpage
\widetext

\begin{center}
\textbf{\large Electronic Supplementary Information for: \\"Promising photovoltaic efficiency of a layered silicon oxide crystal Si\texorpdfstring{\textsubscript{3}}{3}O"} \\
\vspace{10pt}
Sejoong Kim,$^{1,2}$, Kisung Chae,$^{2,3,4, \ast}$ and Young-Woo Son$^{2, \dagger}$ \\
\vspace{4pt}
$^1$ \emph{University of Science and Technology (UST), Daejeon 34113, Korea} \\
$^2$ \emph{Korea Institute for Advanced Study, Seoul 02455, Korea} \\
$^3$ \emph{Department of Chemistry and Biochemistry, University of California, San Diego,La Jolla, CA 92093, United States} \\
$^4$ \emph{Materials Science and Engineering Department, The University of Texas at Dallas, Richardson, Texas 75080, United States} \\
(Dated: \today)
\end{center}

\renewcommand{\thesection}{S\Roman{section}}
\setcounter{section}{0}
\renewcommand{\thefigure}{S\arabic{figure}}
\setcounter{figure}{0}
\renewcommand{\thetable}{S\arabic{table}}
\setcounter{table}{0}
\renewcommand{\bibnumfmt}[1]{[S#1]}
\renewcommand{\citenumfont}[1]{S#1}

\section{Computational Details}
We perform the DFT calculations in order to construct mean-field wavefunctions and energy bands for the $GW$ calculations. 
We use the \textsc{Quantum  Espresso}~\cite{JPhys_CM_21_395502_2009-2,JPhys_CM_29_465901_2017-2} with the plane-wave basis, the PBE exchange-correlation functional~\cite{PhysRevLett_77_3865_1996-2} and norm-conserving pseudopotentials~\cite{PhysRevLett_43_1494-1497_1979-2,PhysRevB_26_4199-4228_1982-2}. 
For the self-consistent calculation, $24\times24\times1$ and $24\times24\times4$ $k$-point grids are adopted for monolayer and bulk respectively. Energy cutoff $952$ eV is used for the plane wave expansion. 
% Figure~\ref{config} shows atomic structures of monolayer and bulk $\textrm{Si}_{3}\textrm{O}$. 
We use the semiemprical Grimme's DFT-D2 scheme~\cite{J_Comp_Chem_27_1787_2006-2} for the van der Waal's correction in order to obtain the fully relaxed structure of the layered bulk Si$_{3}$O.  

The $GW$ calculations are performed by using the \textsc{BerkeleyGW} package~\cite{Com_Phys_Comm_183_1269_2012-2} at the level of $G_{0}W_{0}$ and $GW_{0}$. 
Electronic self-energy is calculated by using the generalized plasmon-pole model~\cite{PhysRevB_34_5390_1986-2} and the modified static remainder approach~\cite{PhysRevB_87_165124_2013-2}. 
The convergence of the quasi-particle (QP) band structure are achieved by tuning parameters such as the $k$-point grid, the energy cutoff of the dielectric matrix $\epsilon_{\mathbf{G},\mathbf{G}^\prime}^{-1}$ and the number of unoccupied bands $N_{b}$. 
The Coulomb interaction truncation scheme~\cite{PhysRevB_73_233103_2006-2} is used to simulate the isolated monolayer geometry of Si$_{3}$O.  
Considering that the QP band structure for low-dimensional systems can show a very slow convergence as a function of the size of the vacuum region as reported in  Ref.~\onlinecite{PhysRevLett_111_216805_2013-2}, the convergence of the QP band structure is also checked by varying the size of vacuum. 
The parameters mentioned above are tuned in order to converge the QP energy gap within $50$ meV. 

\begin{figure}[b]
\begin{center}
\includegraphics[width=0.8\columnwidth, clip=true]{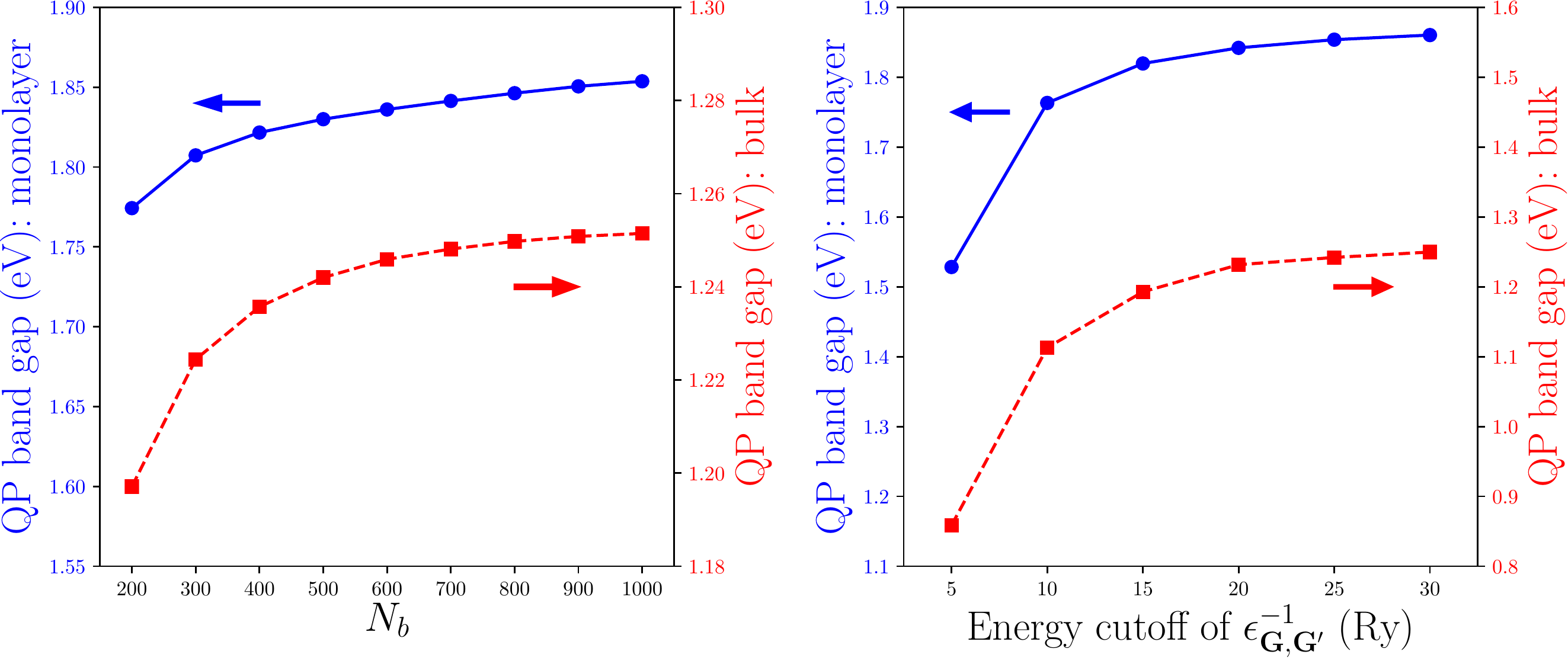} 
\end{center}
\caption{\label{fig:conv} (Color online) Convergence behaviors of QP band gaps as a function of (a) the number of unoccupied bands $N_{b}$ and (b) the energy cutoff of the dielectric function $\epsilon_{\mathbf{G},\mathbf{G}^\prime}^{-1}$. QP band gaps of monolayer and bulk Si$_{3}$O are indicated by blue solid lines and red dashed lines, respectively.}
\end{figure}

We use the energy cutoff $340$ eV for the dielectric matrix $\epsilon_{\mathbf{G},\mathbf{G}^\prime}^{-1}$, and $N_{b}=1000$ unoccupied bands for mono-layer and bulk of $\textrm{Si}_{3}\textrm{O}$. 
Figure~\ref{fig:conv} shows the convergence behavior of the QP band gap at the symmetric point $\textrm{Y}$ as a function of the number of unoccupied bands $N_{b}$ and the energy cutoff of the dielectric function $\epsilon_{\mathbf{G},\mathbf{G}^\prime}^{-1}$. 
In our calculation $14\times14\times1$ and $6\times6\times3$ $k$-grids are sampled for monolayer and bulk, respectively. 
We increase the cell size up to $40\textrm{ \AA}$ along the normal direction to the layer plane. 

%%%%%%%%%%%%%%%%%%%%%%%%%%%%%%%%%%%%%%%%%%%%%%%%%%%%%%%%%%%%%%%%%%%%%%%%%%%%%
% Figure: Convergence test
% (a) QP band gap as a function of N_{b}
% (b) QP band gap as energy cutoff of dielectric matrix
%%%%%%%%%%%%%%%%%%%%%%%%%%%%%%%%%%%%%%%%%%%%%%%%%%%%%%%%%%%%%%%%%%%%%%%%%%%%%
%%%%%%%%%%%%%%%%%%%%%%%%%%%%%%%%%%%%%%%%%%%%%%%%%%%%%%%%%%%%%%%%%%%%%%%%%%%%%

Self-consistently iterative calculations on $G$ are performed to calculate QP band gaps in the so-called $GW_0$ approximation. It is shown that the corresponding band gaps determined by $GW_0$ approximation are converged to the QP band gap linearly interpolated from the $G_0W_0$ gap~\cite{PhysRevB_34_5390_1986-2}. Within three iterations the difference between the $G_{3}W_0$ gap and the linearly interpolated gap becomes smaller than $5$ meV. 

We solve the Bethe-Salpeter equation (BSE) with QP energy bands of Si${_{3}}$O in order to obtain optical absorption spectrum $\epsilon_{2}(\omega)$ (the imaginary part of the dielectric function) and exciton energy levels. 
The numerical solution of the BSE depends on the size of $k$-point mesh and the number of valence and conduction bands. 
$40\times40\times1$ and $20\times20\times4$ $k$-point grids are used to reproduce well-converged absorption spectra for monolayer and bulk $\textrm{Si}_{3}\textrm{O}$, respectively. 
In our calculation six highest valence bands and six lowest conduction ones are used to solve the BSE, and it is shown that the absorption spectrum is well converged up to about $4.0$ eV. Gaussian broadening of $0.05$ eV is adopted to numerically calculate the absorption spectrum. The absorption spectrum is calculated on the energy grid whose interval is $\hbar\Delta\omega=0.01$, on which the numerical integration is performed for the spectroscopic limited maximum efficiency (SLME)~\cite{PhysRevLett_108_068701_2012-2}. 

\begin{figure}[h!]
\begin{center}
\includegraphics[width=0.8\columnwidth]{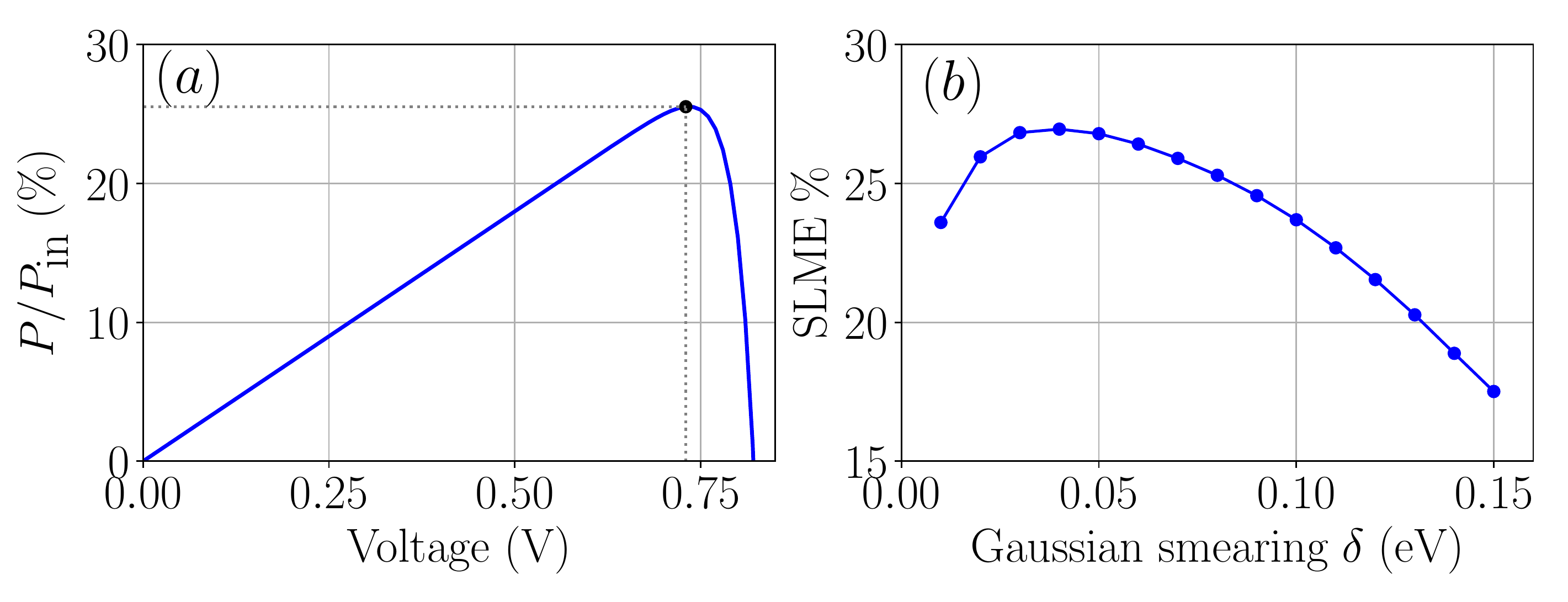} 
\caption{(Color online) (a) $P/P_{in}$ (blue solid line) as a function of the voltage $V$. The SLME $\eta$, the maximum value of $P/P_{in}$, is indicated by black circle. (b) SLME $\eta$ calculations by tuning the broadening parameter from $0.01$ to $0.15$.} 
\label{fig:SLME}
\end{center}
\end{figure}

The analytic expression of the absorption spectrum $\epsilon_{2}(\omega)$ involves the delta function, which can be replaced by the Gaussian function with the broadening parameter in the numerical calculation. 
The broadening parameter, which is in principle small, is needed to be a finite value suitable to numerical integrations for $\epsilon_{2}(\omega)$ and the SLME $\eta$, which are based on the $\mathbf{k}$-point mesh and the discrete grid of $\omega$. 
If the broadening parameter is smaller than the energy resolution of the integration, the spectrum $\epsilon_{2}(\omega)$ shows spurious and bumpy features due to the finite sampling. 
In contrast, too large smearing parameter can wash out important detailed features of the spectrum. We have tested the effect of broadening parameters for $\epsilon_{2}(\omega)$ on SLME calculations. 
The calculations show that the SLME $\eta$ maintains about 26--27\% within the range from 0.02 to 0.06. If the smearing  parameter is  smaller than  the  energy resolution $\hbar\Delta\omega=0.01$ eV, or it is much larger ($> 0.08$ eV), the SLME $\eta$ deviates from 26--27\% as shown in Fig~\ref{fig:SLME}(b).

%%%%%%%%%%%%%%%%%%%%%%%%%%%%%%%%%%%%%%%%%%%%%%%%%%%%%%%%%%%%%%%%%%%%%%%%%%%%%
% Numerical Solution of SLME
%%%%%%%%%%%%%%%%%%%%%%%%%%%%%%%%%%%%%%%%%%%%%%%%%%%%%%%%%%%%%%%%%%%%%%%%%%%%%
%%%%%%%%%%%%%%%%%%%%%%%%%%%%%%%%%%%%%%%%%%%%%%%%%%%%%%%%%%%%%%%%%%%%%%%%%%%%%

\section{SLME Calculations}
The SLME $\eta$ of the solar cell can be calculated by maximizing the ratio $P/P_{in}$~\cite{PhysRevLett_108_068701_2012-2,acs.chemmater.9b02166-2}. Here the total incident solar energy density $P_{in}$ is calculated by using the Air Mass 1.5 data~\cite{AM1.5-2} for the solar irradiance spectrum of the photon flux $I_{AM1.5}(E)$, 
\begin{equation}
P_{in} = \int_{0}^{\infty} I_{AM1.5}(E) E dE.
\end{equation}
The output power density of the solar cell $P$ is the product of the total net current density $J$ and the voltage $V$, 
\begin{equation}
P = JV = \left[J_{sc} - J_{0}\left(e^{eV/k_{B}T} - 1 \right) \right]V,    
\end{equation}
where $e$, $k_{B}$, and $T$ are the electron charge, the Boltzmann constant, and the solar cell temperature, respectively~\cite{PhysRevLett_108_068701_2012-2,acs.chemmater.9b02166-2}. The net current density $J$ is determined by two contributions: the short-circuit current density $J_{sc}$ and the reverse saturation current density $J_{0}$. The short-circuit current density $J_{sc}$ is calculated from the absorbance $A(E)$ and the AM1.5 spectrum $I_{AM1.5}(E)$, 
\begin{equation}
J_{sc} = e\int_{0}^{\infty}A(E) I_{AM1.5}(E)dE.
\end{equation}
$J_{0}$ is further expressed as the sum of the non-radiative electron-hole combination current density $J_{0}^{nr}$ and the radiative one $J_{0}^{r}$, 
\begin{equation}
J_{0} = J_{0}^{nr} + J_{0}^{r} = \frac{J_{0}^{r}}{f_{r}}.
\end{equation}
Here $f_{r}=J_{0}^{r}/\left(J_{0}^{nr} + J_{0}^{r}\right)$ is the fraction of the radiative combination current, which is approximately given by $f_{r} = e^{-(E_{g}^{da}-E_{g})/k_{B}T}$, where $E_{g}$ and $E_{g}^{da}$ are the minimum gap and the directly allowed gap, respectively~\cite{PhysRevLett_108_068701_2012-2,acs.chemmater.9b02166-2}.
Considering the principle of the detailed balance, the radiative combination current $J_{0}^{r}$ is equal to the absorption rate of photons from the surrounding thermal path in equilibrium with the solar cell surface:
\begin{equation}
J_{0}^{r} = e\pi \int^{\infty}_{0} A(E) I_{bb}(E,T)dE,
\end{equation}
where $I_{bb}(E,T)$ stands for the spectrum of the black body at temperature $T$, 
\begin{equation}
I_{bb}(E,T) = \frac{2\pi}{h^3 c^2}\frac{E^2}{e^{E/k_{B}T}-1},
\end{equation}
where $h$ and $c$ are the Planck constant and the speed of light, and the temperature of the surrounding thermal bath is $T=25^{\circ}\textrm{C}$ in this work. The SLME $\eta$ can be obtained by numerically maximizing $P$ as shown in Fig.~\ref{fig:SLME}(a).

\section{Thermal Stability of \texorpdfstring{$\textrm{Si}_{3}\textrm{O}$}{TEXT}}

%%%%%%%%%%%%%%%%%%%%%%%%%%%%%%%%%%%%%%%%%%%%%%%%%%%%%%%%%%%%%%%%%%%%%%%%%%%%%
% Thermal Stability of AIMD
%%%%%%%%%%%%%%%%%%%%%%%%%%%%%%%%%%%%%%%%%%%%%%%%%%%%%%%%%%%%%%%%%%%%%%%%%%%%%
\begin{figure}[b!]
\begin{center}
\includegraphics[width=0.6\columnwidth]{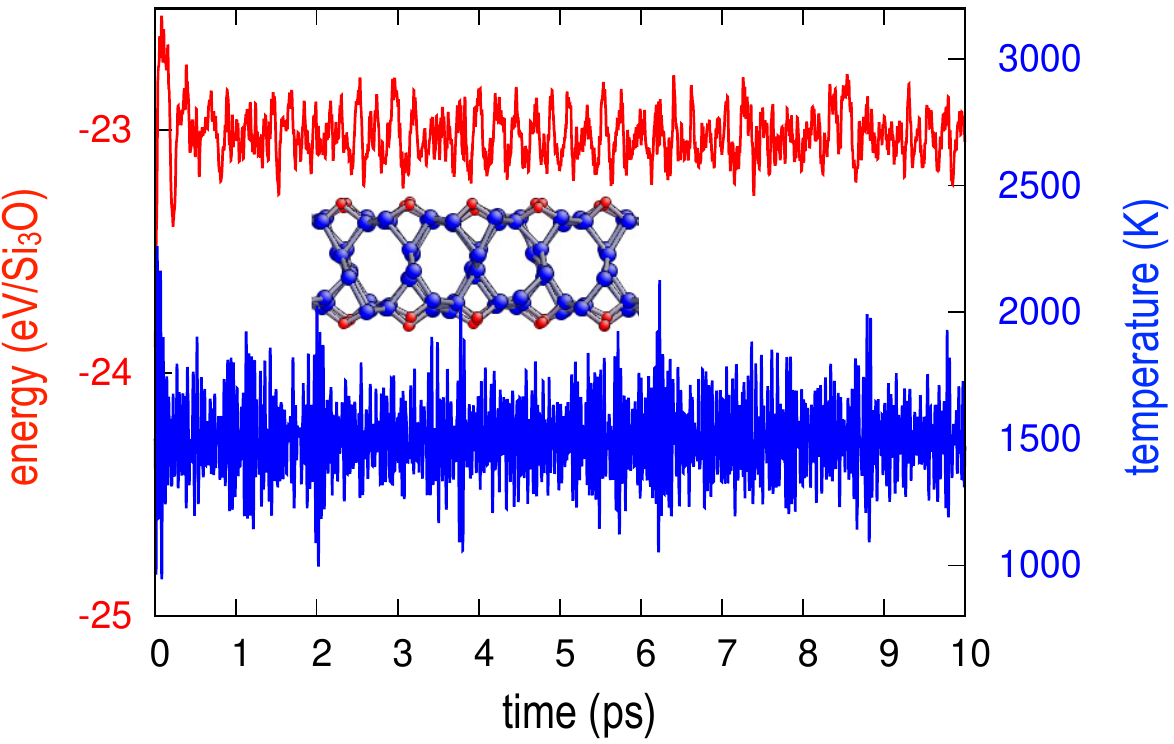} 
\caption{(Color online) Evolution of energy (red) and temperature (blue) with elapsed time. Snapshot of Si$_{3}$O is shown in the inset.}
\label{fig:AIMD}
\end{center}
\end{figure}
%%%%%%%%%%%%%%%%%%%%%%%%%%%%%%%%%%%%%%%%%%%%%%%%%%%%%%%%%%%%%%%%%%%%%%%%%%%%%

Here, \emph{ab initio} molecular dynamics (AIMD) is employed to show robust thermal stability of Si$_{3}$O in addition to convex Hull and harmonic phonon dispersion provided in Ref.~\onlinecite{NanoLett_19_2694_2019-2}.
The Si$_{3}$O monolayer is expanded to a (3$\times$3$\times$1) supercell which contains 18 Si$_{3}$O formula units.
AIMD is performed at a temperature of 1,500 K in canonical ensemble (i.e., constant NVT), where temperature of the system is controlled by Nos\'e-Hoover thermostat.
The system is integrated by using Verlet algorithm for 10 pico seconds (or 10,000 steps) with a time step of 1 femto second.
Fig.~\ref{fig:AIMD} shows temporal evolution of energy and temperature of the system during the AIMD calculation.
The instantaneous energy seems to fluctuate around a constant value within a reasonable energy window, indicating that the temporal average is kept constant.
This means that the atomic arrangements in the Si$_{3}$O remain stable without any broken bonds throughout the AIMD calculations.
The inset in Fig.~\ref{fig:AIMD} shows a snapshot of the Si$_{3}$O monolayer during the AIMD.
Each of the atoms vibrates at their equilibrium positions due to the kinetic energy, but the initial structure is maintained, confirming the robust thermal stability.

\section{Effective masses in \texorpdfstring{$\textrm{Si}_{3}\textrm{O}$}{TEXT}}
%%%%%%%%%%%%%%%%%%%%%%%%%%%%%%%%%%%%%%%%%%%%%%%%%%%%%%%%%%%%%%%%%%%%%%%%%%%%%
% Effective masses in Si3O
%%%%%%%%%%%%%%%%%%%%%%%%%%%%%%%%%%%%%%%%%%%%%%%%%%%%%%%%%%%%%%%%%%%%%%%%%%%%%
\begin{figure}[h!]
\begin{center}
\includegraphics[width=0.8\columnwidth]{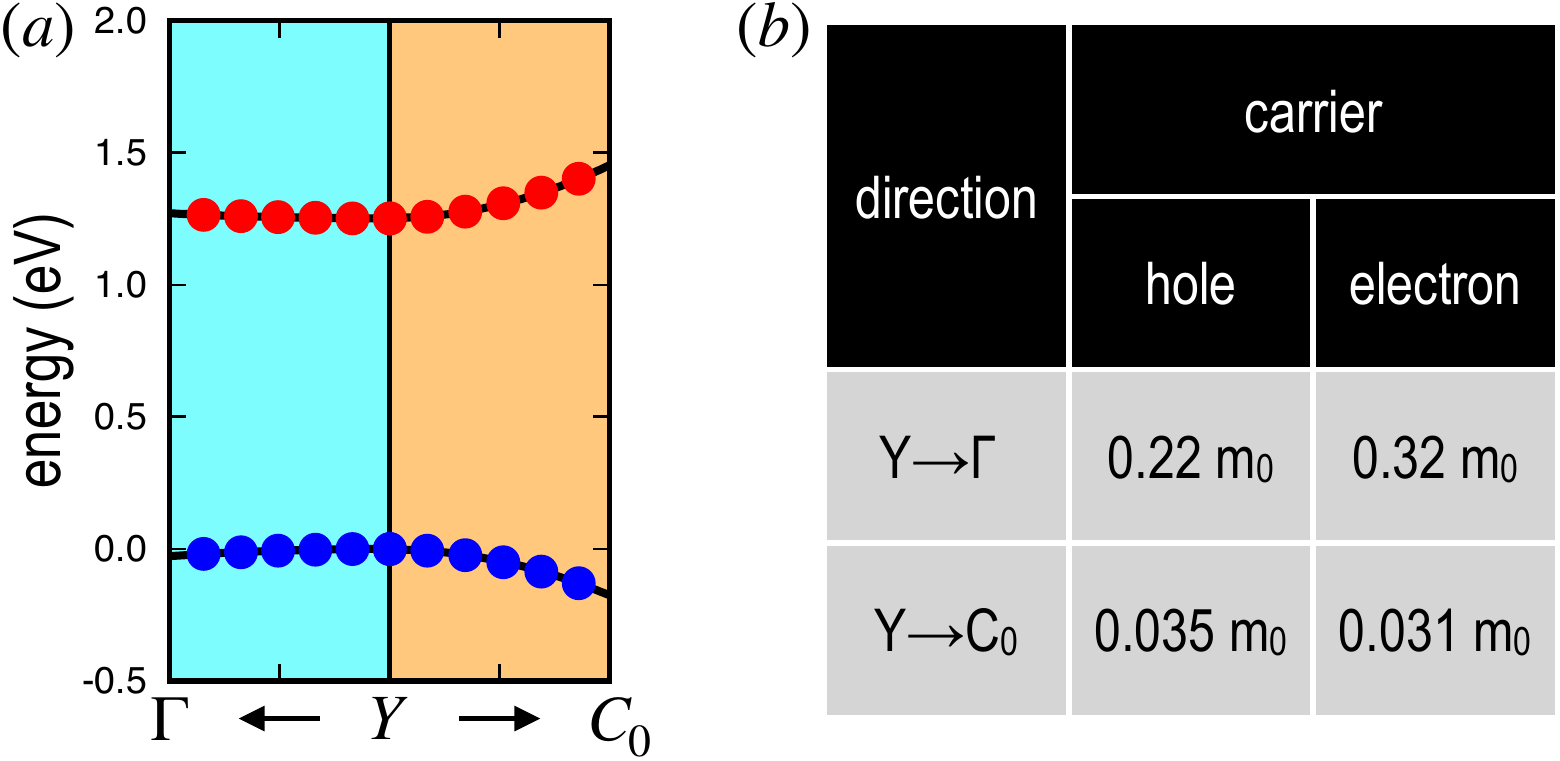} 
\caption{(Color online) Effective masses of Si$_{3}$O. (a) Quasiparticle eigenvalues near the band edges: conduction (red symbols) and valence (blue symbols) bands along $Y$-$\Gamma$ (cyan shading) and $Y$-$C_{0}$ (orange shading) directions. The black curves behind the data points are fitted to quadratic functions. (b) The calculated effective masses are tabulated relative to an electron rest mass $m_{0}$.}
\label{fig:effmass}
\end{center}
\end{figure}
%%%%%%%%%%%%%%%%%%%%%%%%%%%%%%%%%%%%%%%%%%%%%%%%%%%%%%%%%%%%%%%%%%%%%%%%%%%%%
Carrier mobility ($\mu$) is an important characteristic in photovoltaic applications for efficient charge separation and photo-current collection.
In solids, the carrier mobility can be shown as 
$$\mu=\frac{q\tau}{2m^{*}}$$
where $q$ is an elemental charge, $\tau$ is scattering time and $m^{*}$ is an effective mass. 
The scattering time $\tau$ depends on various factors such as details of electronic structures, defect concentrations and temperature, and electron-phonon scattering becomes a dominant factor for pure bulk materials with few defects.
While elaborated evaluation of the $\tau$ is crucial for the quantitatively assessment of carrier mobility, it will require demanding computation of electron-phonon coupling matrix, which is beyond the scope of this paper.
As a crude approximation, we hypothesize that effective masses ($m^{*}$) be sufficient to possibly show qualitative picture of carrier mobility behaviors in Si$_{3}$O.
We investigate the in-plane effective masses in bulk Si$_{3}$O from quasiparticle band structures as shown in Fig. 2(b).
Two high-symmetry paths are considered: from $Y$ to $\Gamma$ and from $Y$ to $C_{0}$, and small fractions of each path around the $Y$ momentum, where both maximum and minimum occur, are shown in Fig.~\ref{fig:effmass}(a).
Quasiparticle eigenvalues along each of the paths in each band are used to separately fit the harmonic energy-momentum (E-k) dispersion behavior near the band edges, i.e.,
$$E(k)=E_{0}+\frac{\hbar k^{2}}{2m^{*}}$$
where $\hbar$ is Planck constant.
The effective mass can be obtained from the curvature of the second derivative of the band structure,
$$\left(m^{*}\right)^{-1}=\frac{1}{\hbar}\frac{d^2E}{dk^2}$$
which is usually used relative to the electron rest mass $m_{0}$.
Figure~\ref{fig:effmass}(b) shows the $m^{*}$ for each direction and for both electron and hole.
It is interesting that the effective masses for both electron and hole along different directions show an order-of-magnitude difference, indicating that the Si$_{3}$O is highly anisotropic.
Moreover, the effective mass values of the \emph{light} bands for both electron and hole are remarkably small ($\sim$0.03 $m_{0}$), comparable to well-known high mobility semiconductors such as InSb (0.0135 $m_{0}$) and GaAs (0.067 $m_{0}$)~\cite{doi:10.1063/1.1368156-2}.
Even for the \emph{heavy} bands, the effective mass values are comparable to a bulk silicon crystal (0.19 $m_{0}$ and 0.16 $m_{0}$ for electron and hole, respectively).
It is worth noting that the band edges at $Y$ serve as the sole predominant inter-band transition path across the band gap up to a few hundreds meV, limiting the number of possible electron-phonon coupling pathways up to considerable temperature range.
With remarkably small effective masses for both electron and hole as well as the interesting electronic structures favorable for efficient carrier transport, Si$_{3}$O makes a promising candidate for photovoltaic applications.

%\bibliography{bibl}

\begin{thebibliography}{47}%
\makeatletter
\providecommand \@ifxundefined [1]{%
 \@ifx{#1\undefined}
}%
\providecommand \@ifnum [1]{%
 \ifnum #1\expandafter \@firstoftwo
 \else \expandafter \@secondoftwo
 \fi
}%
\providecommand \@ifx [1]{%
 \ifx #1\expandafter \@firstoftwo
 \else \expandafter \@secondoftwo
 \fi
}%
\providecommand \natexlab [1]{#1}%
\providecommand \enquote  [1]{``#1''}%
\providecommand \bibnamefont  [1]{#1}%
\providecommand \bibfnamefont [1]{#1}%
\providecommand \citenamefont [1]{#1}%
\providecommand \href@noop [0]{\@secondoftwo}%
\providecommand \href [0]{\begingroup \@sanitize@url \@href}%
\providecommand \@href[1]{\@@startlink{#1}\@@href}%
\providecommand \@@href[1]{\endgroup#1\@@endlink}%
\providecommand \@sanitize@url [0]{\catcode `\\12\catcode `\$12\catcode
  `\&12\catcode `\#12\catcode `\^12\catcode `\_12\catcode `\%12\relax}%
\providecommand \@@startlink[1]{}%
\providecommand \@@endlink[0]{}%
\providecommand \url  [0]{\begingroup\@sanitize@url \@url }%
\providecommand \@url [1]{\endgroup\@href {#1}{\urlprefix }}%
\providecommand \urlprefix  [0]{URL }%
\providecommand \Eprint [0]{\href }%
\providecommand \doibase [0]{http://dx.doi.org/}%
\providecommand \selectlanguage [0]{\@gobble}%
\providecommand \bibinfo  [0]{\@secondoftwo}%
\providecommand \bibfield  [0]{\@secondoftwo}%
\providecommand \translation [1]{[#1]}%
\providecommand \BibitemOpen [0]{}%
\providecommand \bibitemStop [0]{}%
\providecommand \bibitemNoStop [0]{.\EOS\space}%
\providecommand \EOS [0]{\spacefactor3000\relax}%
\providecommand \BibitemShut  [1]{\csname bibitem#1\endcsname}%
\let\auto@bib@innerbib\@empty
%</preamble>
\bibitem [{\citenamefont {Haegel}\ \emph {et~al.}(2017)\citenamefont {Haegel},
  \citenamefont {Margolis}, \citenamefont {Buonassisi}, \citenamefont
  {Feldman}, \citenamefont {Froitzheim}, \citenamefont {Garabedian},
  \citenamefont {Green}, \citenamefont {Glunz}, \citenamefont {Henning},
  \citenamefont {Holder}, \citenamefont {Kaizuka}, \citenamefont {Kroposki},
  \citenamefont {Matsubara}, \citenamefont {Niki}, \citenamefont {Sakurai},
  \citenamefont {Schindler}, \citenamefont {Tumas}, \citenamefont {Weber},
  \citenamefont {Wilson}, \citenamefont {Woodhouse},\ and\ \citenamefont
  {Kurtz}}]{Haegel141}%
  \BibitemOpen
  \bibfield  {author} {\bibinfo {author} {\bibfnamefont {N.~M.}\ \bibnamefont
  {Haegel}}, \bibinfo {author} {\bibfnamefont {R.}~\bibnamefont {Margolis}},
  \bibinfo {author} {\bibfnamefont {T.}~\bibnamefont {Buonassisi}}, \bibinfo
  {author} {\bibfnamefont {D.}~\bibnamefont {Feldman}}, \bibinfo {author}
  {\bibfnamefont {A.}~\bibnamefont {Froitzheim}}, \bibinfo {author}
  {\bibfnamefont {R.}~\bibnamefont {Garabedian}}, \bibinfo {author}
  {\bibfnamefont {M.}~\bibnamefont {Green}}, \bibinfo {author} {\bibfnamefont
  {S.}~\bibnamefont {Glunz}}, \bibinfo {author} {\bibfnamefont {H.-M.}\
  \bibnamefont {Henning}}, \bibinfo {author} {\bibfnamefont {B.}~\bibnamefont
  {Holder}}, \bibinfo {author} {\bibfnamefont {I.}~\bibnamefont {Kaizuka}},
  \bibinfo {author} {\bibfnamefont {B.}~\bibnamefont {Kroposki}}, \bibinfo
  {author} {\bibfnamefont {K.}~\bibnamefont {Matsubara}}, \bibinfo {author}
  {\bibfnamefont {S.}~\bibnamefont {Niki}}, \bibinfo {author} {\bibfnamefont
  {K.}~\bibnamefont {Sakurai}}, \bibinfo {author} {\bibfnamefont {R.~A.}\
  \bibnamefont {Schindler}}, \bibinfo {author} {\bibfnamefont {W.}~\bibnamefont
  {Tumas}}, \bibinfo {author} {\bibfnamefont {E.~R.}\ \bibnamefont {Weber}},
  \bibinfo {author} {\bibfnamefont {G.}~\bibnamefont {Wilson}}, \bibinfo
  {author} {\bibfnamefont {M.}~\bibnamefont {Woodhouse}}, \ and\ \bibinfo
  {author} {\bibfnamefont {S.}~\bibnamefont {Kurtz}},\ }\href@noop {}
  {\bibfield  {journal} {\bibinfo  {journal} {Science}\ }\textbf {\bibinfo
  {volume} {356}},\ \bibinfo {pages} {141} (\bibinfo {year}
  {2017})}\BibitemShut {NoStop}%
\bibitem [{\citenamefont {Wang}\ \emph {et~al.}(2012)\citenamefont {Wang},
  \citenamefont {Kalantar-Zadeh}, \citenamefont {Kis}, \citenamefont
  {Coleman},\ and\ \citenamefont {Strano}}]{StranoNatureNano}%
  \BibitemOpen
  \bibfield  {author} {\bibinfo {author} {\bibfnamefont {Q.~H.}\ \bibnamefont
  {Wang}}, \bibinfo {author} {\bibfnamefont {K.}~\bibnamefont
  {Kalantar-Zadeh}}, \bibinfo {author} {\bibfnamefont {A.}~\bibnamefont {Kis}},
  \bibinfo {author} {\bibfnamefont {J.~N.}\ \bibnamefont {Coleman}}, \ and\
  \bibinfo {author} {\bibfnamefont {M.~S.}\ \bibnamefont {Strano}},\
  }\href@noop {} {\bibfield  {journal} {\bibinfo  {journal} {Nat. Nano.}\
  }\textbf {\bibinfo {volume} {7}},\ \bibinfo {pages} {699} (\bibinfo {year}
  {2012})}\BibitemShut {NoStop}%
\bibitem [{\citenamefont {Ferrari}\ \emph {et~al.}(2015)\citenamefont
  {Ferrari}, \citenamefont {Bonaccorso}, \citenamefont {Fal{'}ko},
  \citenamefont {Novoselov}, \citenamefont {Roche}, \citenamefont {B{\o}ggild},
  \citenamefont {Borini}, \citenamefont {Koppens}, \citenamefont {Palermo},
  \citenamefont {Pugno}, \citenamefont {Garrido}, \citenamefont {Sordan},
  \citenamefont {Bianco}, \citenamefont {Ballerini}, \citenamefont {Prato},
  \citenamefont {Lidorikis}, \citenamefont {Kivioja}, \citenamefont
  {Marinelli}, \citenamefont {Ryh{\"a}nen}, \citenamefont {Morpurgo},
  \citenamefont {Coleman}, \citenamefont {Nicolosi}, \citenamefont {Colombo},
  \citenamefont {Fert}, \citenamefont {Garcia-Hernandez}, \citenamefont
  {Bachtold}, \citenamefont {Schneider}, \citenamefont {Guinea}, \citenamefont
  {Dekker}, \citenamefont {Barbone}, \citenamefont {Sun}, \citenamefont
  {Galiotis}, \citenamefont {Grigorenko}, \citenamefont {Konstantatos},
  \citenamefont {Kis}, \citenamefont {Katsnelson}, \citenamefont {Vandersypen},
  \citenamefont {Loiseau}, \citenamefont {Morandi}, \citenamefont {Neumaier},
  \citenamefont {Treossi}, \citenamefont {Pellegrini}, \citenamefont {Polini},
  \citenamefont {Tredicucci}, \citenamefont {Williams}, \citenamefont
  {Hee~Hong}, \citenamefont {Ahn}, \citenamefont {Min~Kim}, \citenamefont
  {Zirath}, \citenamefont {van Wees}, \citenamefont {van~der Zant},
  \citenamefont {Occhipinti}, \citenamefont {Di~Matteo}, \citenamefont
  {Kinloch}, \citenamefont {Seyller}, \citenamefont {Quesnel}, \citenamefont
  {Feng}, \citenamefont {Teo}, \citenamefont {Rupesinghe}, \citenamefont
  {Hakonen}, \citenamefont {Neil}, \citenamefont {Tannock}, \citenamefont
  {L{\"o}fwander},\ and\ \citenamefont {Kinaret}}]{C4NR01600A}%
  \BibitemOpen
  \bibfield  {author} {\bibinfo {author} {\bibfnamefont {A.~C.}\ \bibnamefont
  {Ferrari}}, \bibinfo {author} {\bibfnamefont {F.}~\bibnamefont {Bonaccorso}},
  \bibinfo {author} {\bibfnamefont {V.}~\bibnamefont {Fal{'}ko}}, \bibinfo
  {author} {\bibfnamefont {K.~S.}\ \bibnamefont {Novoselov}}, \bibinfo {author}
  {\bibfnamefont {S.}~\bibnamefont {Roche}}, \bibinfo {author} {\bibfnamefont
  {P.}~\bibnamefont {B{\o}ggild}}, \bibinfo {author} {\bibfnamefont
  {S.}~\bibnamefont {Borini}}, \bibinfo {author} {\bibfnamefont {F.~H.~L.}\
  \bibnamefont {Koppens}}, \bibinfo {author} {\bibfnamefont {V.}~\bibnamefont
  {Palermo}}, \bibinfo {author} {\bibfnamefont {N.}~\bibnamefont {Pugno}},
  \bibinfo {author} {\bibfnamefont {J.~A.}\ \bibnamefont {Garrido}}, \bibinfo
  {author} {\bibfnamefont {R.}~\bibnamefont {Sordan}}, \bibinfo {author}
  {\bibfnamefont {A.}~\bibnamefont {Bianco}}, \bibinfo {author} {\bibfnamefont
  {L.}~\bibnamefont {Ballerini}}, \bibinfo {author} {\bibfnamefont
  {M.}~\bibnamefont {Prato}}, \bibinfo {author} {\bibfnamefont
  {E.}~\bibnamefont {Lidorikis}}, \bibinfo {author} {\bibfnamefont
  {J.}~\bibnamefont {Kivioja}}, \bibinfo {author} {\bibfnamefont
  {C.}~\bibnamefont {Marinelli}}, \bibinfo {author} {\bibfnamefont
  {T.}~\bibnamefont {Ryh{\"a}nen}}, \bibinfo {author} {\bibfnamefont
  {A.}~\bibnamefont {Morpurgo}}, \bibinfo {author} {\bibfnamefont {J.~N.}\
  \bibnamefont {Coleman}}, \bibinfo {author} {\bibfnamefont {V.}~\bibnamefont
  {Nicolosi}}, \bibinfo {author} {\bibfnamefont {L.}~\bibnamefont {Colombo}},
  \bibinfo {author} {\bibfnamefont {A.}~\bibnamefont {Fert}}, \bibinfo {author}
  {\bibfnamefont {M.}~\bibnamefont {Garcia-Hernandez}}, \bibinfo {author}
  {\bibfnamefont {A.}~\bibnamefont {Bachtold}}, \bibinfo {author}
  {\bibfnamefont {G.~F.}\ \bibnamefont {Schneider}}, \bibinfo {author}
  {\bibfnamefont {F.}~\bibnamefont {Guinea}}, \bibinfo {author} {\bibfnamefont
  {C.}~\bibnamefont {Dekker}}, \bibinfo {author} {\bibfnamefont
  {M.}~\bibnamefont {Barbone}}, \bibinfo {author} {\bibfnamefont
  {Z.}~\bibnamefont {Sun}}, \bibinfo {author} {\bibfnamefont {C.}~\bibnamefont
  {Galiotis}}, \bibinfo {author} {\bibfnamefont {A.~N.}\ \bibnamefont
  {Grigorenko}}, \bibinfo {author} {\bibfnamefont {G.}~\bibnamefont
  {Konstantatos}}, \bibinfo {author} {\bibfnamefont {A.}~\bibnamefont {Kis}},
  \bibinfo {author} {\bibfnamefont {M.}~\bibnamefont {Katsnelson}}, \bibinfo
  {author} {\bibfnamefont {L.}~\bibnamefont {Vandersypen}}, \bibinfo {author}
  {\bibfnamefont {A.}~\bibnamefont {Loiseau}}, \bibinfo {author} {\bibfnamefont
  {V.}~\bibnamefont {Morandi}}, \bibinfo {author} {\bibfnamefont
  {D.}~\bibnamefont {Neumaier}}, \bibinfo {author} {\bibfnamefont
  {E.}~\bibnamefont {Treossi}}, \bibinfo {author} {\bibfnamefont
  {V.}~\bibnamefont {Pellegrini}}, \bibinfo {author} {\bibfnamefont
  {M.}~\bibnamefont {Polini}}, \bibinfo {author} {\bibfnamefont
  {A.}~\bibnamefont {Tredicucci}}, \bibinfo {author} {\bibfnamefont {G.~M.}\
  \bibnamefont {Williams}}, \bibinfo {author} {\bibfnamefont {B.}~\bibnamefont
  {Hee~Hong}}, \bibinfo {author} {\bibfnamefont {J.-H.}\ \bibnamefont {Ahn}},
  \bibinfo {author} {\bibfnamefont {J.}~\bibnamefont {Min~Kim}}, \bibinfo
  {author} {\bibfnamefont {H.}~\bibnamefont {Zirath}}, \bibinfo {author}
  {\bibfnamefont {B.~J.}\ \bibnamefont {van Wees}}, \bibinfo {author}
  {\bibfnamefont {H.}~\bibnamefont {van~der Zant}}, \bibinfo {author}
  {\bibfnamefont {L.}~\bibnamefont {Occhipinti}}, \bibinfo {author}
  {\bibfnamefont {A.}~\bibnamefont {Di~Matteo}}, \bibinfo {author}
  {\bibfnamefont {I.~A.}\ \bibnamefont {Kinloch}}, \bibinfo {author}
  {\bibfnamefont {T.}~\bibnamefont {Seyller}}, \bibinfo {author} {\bibfnamefont
  {E.}~\bibnamefont {Quesnel}}, \bibinfo {author} {\bibfnamefont
  {X.}~\bibnamefont {Feng}}, \bibinfo {author} {\bibfnamefont {K.}~\bibnamefont
  {Teo}}, \bibinfo {author} {\bibfnamefont {N.}~\bibnamefont {Rupesinghe}},
  \bibinfo {author} {\bibfnamefont {P.}~\bibnamefont {Hakonen}}, \bibinfo
  {author} {\bibfnamefont {S.~R.~T.}\ \bibnamefont {Neil}}, \bibinfo {author}
  {\bibfnamefont {Q.}~\bibnamefont {Tannock}}, \bibinfo {author} {\bibfnamefont
  {T.}~\bibnamefont {L{\"o}fwander}}, \ and\ \bibinfo {author} {\bibfnamefont
  {J.}~\bibnamefont {Kinaret}},\ }\href@noop {} {\bibfield  {journal} {\bibinfo
   {journal} {Nanoscale}\ }\textbf {\bibinfo {volume} {7}},\ \bibinfo {pages}
  {4598} (\bibinfo {year} {2015})}\BibitemShut {NoStop}%
\bibitem [{\citenamefont {Bonaccorso}\ \emph {et~al.}(2015)\citenamefont
  {Bonaccorso}, \citenamefont {Colombo}, \citenamefont {Yu}, \citenamefont
  {Stoller}, \citenamefont {Tozzini}, \citenamefont {Ferrari}, \citenamefont
  {Ruoff},\ and\ \citenamefont {Pellegrini}}]{Bonaccorso1246501}%
  \BibitemOpen
  \bibfield  {author} {\bibinfo {author} {\bibfnamefont {F.}~\bibnamefont
  {Bonaccorso}}, \bibinfo {author} {\bibfnamefont {L.}~\bibnamefont {Colombo}},
  \bibinfo {author} {\bibfnamefont {G.}~\bibnamefont {Yu}}, \bibinfo {author}
  {\bibfnamefont {M.}~\bibnamefont {Stoller}}, \bibinfo {author} {\bibfnamefont
  {V.}~\bibnamefont {Tozzini}}, \bibinfo {author} {\bibfnamefont {A.~C.}\
  \bibnamefont {Ferrari}}, \bibinfo {author} {\bibfnamefont {R.~S.}\
  \bibnamefont {Ruoff}}, \ and\ \bibinfo {author} {\bibfnamefont
  {V.}~\bibnamefont {Pellegrini}},\ }\href@noop {} {\bibfield  {journal}
  {\bibinfo  {journal} {Science}\ }\textbf {\bibinfo {volume} {347}},\ \bibinfo
  {pages} {1246501} (\bibinfo {year} {2015})}\BibitemShut {NoStop}%
\bibitem [{\citenamefont {Novoselov}\ \emph {et~al.}(2016)\citenamefont
  {Novoselov}, \citenamefont {Mishchenko}, \citenamefont {Carvalho},\ and\
  \citenamefont {Castro~Neto}}]{Novoselovaac9439}%
  \BibitemOpen
  \bibfield  {author} {\bibinfo {author} {\bibfnamefont {K.~S.}\ \bibnamefont
  {Novoselov}}, \bibinfo {author} {\bibfnamefont {A.}~\bibnamefont
  {Mishchenko}}, \bibinfo {author} {\bibfnamefont {A.}~\bibnamefont
  {Carvalho}}, \ and\ \bibinfo {author} {\bibfnamefont {A.~H.}\ \bibnamefont
  {Castro~Neto}},\ }\href {\doibase 10.1126/science.aac9439} {\bibfield
  {journal} {\bibinfo  {journal} {Science}\ }\textbf {\bibinfo {volume}
  {353}},\ \bibinfo {pages} {aac9439} (\bibinfo {year} {2016})}\BibitemShut
  {NoStop}%
\bibitem [{\citenamefont {Jariwala}\ \emph {et~al.}(2017)\citenamefont
  {Jariwala}, \citenamefont {Davoyan}, \citenamefont {Wong},\ and\
  \citenamefont {Atwater}}]{acsphotonics.7b01103}%
  \BibitemOpen
  \bibfield  {author} {\bibinfo {author} {\bibfnamefont {D.}~\bibnamefont
  {Jariwala}}, \bibinfo {author} {\bibfnamefont {A.~R.}\ \bibnamefont
  {Davoyan}}, \bibinfo {author} {\bibfnamefont {J.}~\bibnamefont {Wong}}, \
  and\ \bibinfo {author} {\bibfnamefont {H.~A.}\ \bibnamefont {Atwater}},\
  }\href@noop {} {\bibfield  {journal} {\bibinfo  {journal} {ACS Photonics}\
  }\textbf {\bibinfo {volume} {4}},\ \bibinfo {pages} {2962} (\bibinfo {year}
  {2017})}\BibitemShut {NoStop}%
\bibitem [{\citenamefont {Das}\ \emph {et~al.}(2019)\citenamefont {Das},
  \citenamefont {Pandey}, \citenamefont {Thomas},\ and\ \citenamefont
  {Roy}}]{adma.201802722}%
  \BibitemOpen
  \bibfield  {author} {\bibinfo {author} {\bibfnamefont {S.}~\bibnamefont
  {Das}}, \bibinfo {author} {\bibfnamefont {D.}~\bibnamefont {Pandey}},
  \bibinfo {author} {\bibfnamefont {J.}~\bibnamefont {Thomas}}, \ and\ \bibinfo
  {author} {\bibfnamefont {T.}~\bibnamefont {Roy}},\ }\href@noop {} {\bibfield
  {journal} {\bibinfo  {journal} {Adv. Mat.}\ }\textbf {\bibinfo {volume}
  {31}},\ \bibinfo {pages} {1802722} (\bibinfo {year} {2019})}\BibitemShut
  {NoStop}%
\bibitem [{\citenamefont {Ashton}\ \emph {et~al.}(2017)\citenamefont {Ashton},
  \citenamefont {Paul}, \citenamefont {Sinnott},\ and\ \citenamefont
  {Hennig}}]{PhysRevLett.118.106101}%
  \BibitemOpen
  \bibfield  {author} {\bibinfo {author} {\bibfnamefont {M.}~\bibnamefont
  {Ashton}}, \bibinfo {author} {\bibfnamefont {J.}~\bibnamefont {Paul}},
  \bibinfo {author} {\bibfnamefont {S.~B.}\ \bibnamefont {Sinnott}}, \ and\
  \bibinfo {author} {\bibfnamefont {R.~G.}\ \bibnamefont {Hennig}},\
  }\href@noop {} {\bibfield  {journal} {\bibinfo  {journal} {Phys. Rev. Lett.}\
  }\textbf {\bibinfo {volume} {118}},\ \bibinfo {pages} {106101} (\bibinfo
  {year} {2017})}\BibitemShut {NoStop}%
\bibitem [{\citenamefont {Mounet}\ \emph {et~al.}(2018)\citenamefont {Mounet},
  \citenamefont {Gibertini}, \citenamefont {Schwaller}, \citenamefont {Campi},
  \citenamefont {Merkys}, \citenamefont {Marrazzo}, \citenamefont {Sohier},
  \citenamefont {Castelli}, \citenamefont {Cepellotti}, \citenamefont {Pizzi},\
  and\ \citenamefont {Marzari}}]{Marzari2018}%
  \BibitemOpen
  \bibfield  {author} {\bibinfo {author} {\bibfnamefont {N.}~\bibnamefont
  {Mounet}}, \bibinfo {author} {\bibfnamefont {M.}~\bibnamefont {Gibertini}},
  \bibinfo {author} {\bibfnamefont {P.}~\bibnamefont {Schwaller}}, \bibinfo
  {author} {\bibfnamefont {D.}~\bibnamefont {Campi}}, \bibinfo {author}
  {\bibfnamefont {A.}~\bibnamefont {Merkys}}, \bibinfo {author} {\bibfnamefont
  {A.}~\bibnamefont {Marrazzo}}, \bibinfo {author} {\bibfnamefont
  {T.}~\bibnamefont {Sohier}}, \bibinfo {author} {\bibfnamefont {I.~E.}\
  \bibnamefont {Castelli}}, \bibinfo {author} {\bibfnamefont {A.}~\bibnamefont
  {Cepellotti}}, \bibinfo {author} {\bibfnamefont {G.}~\bibnamefont {Pizzi}}, \
  and\ \bibinfo {author} {\bibfnamefont {N.}~\bibnamefont {Marzari}},\
  }\href@noop {} {\bibfield  {journal} {\bibinfo  {journal} {Nature
  Nanotechnology}\ }\textbf {\bibinfo {volume} {13}},\ \bibinfo {pages} {246}
  (\bibinfo {year} {2018})}\BibitemShut {NoStop}%
\bibitem [{\citenamefont {Bush}\ \emph {et~al.}(1995)\citenamefont {Bush},
  \citenamefont {Catlow},\ and\ \citenamefont {Battle}}]{JM9950501269}%
  \BibitemOpen
  \bibfield  {author} {\bibinfo {author} {\bibfnamefont {T.~S.}\ \bibnamefont
  {Bush}}, \bibinfo {author} {\bibfnamefont {C.~R.~A.}\ \bibnamefont {Catlow}},
  \ and\ \bibinfo {author} {\bibfnamefont {P.~D.}\ \bibnamefont {Battle}},\
  }\href@noop {} {\bibfield  {journal} {\bibinfo  {journal} {J. Mater. Chem.}\
  }\textbf {\bibinfo {volume} {5}},\ \bibinfo {pages} {1269} (\bibinfo {year}
  {1995})}\BibitemShut {NoStop}%
\bibitem [{\citenamefont {Oganov}\ and\ \citenamefont {Glass}(2006)}]{Oganov}%
  \BibitemOpen
  \bibfield  {author} {\bibinfo {author} {\bibfnamefont {A.~R.}\ \bibnamefont
  {Oganov}}\ and\ \bibinfo {author} {\bibfnamefont {C.~W.}\ \bibnamefont
  {Glass}},\ }\href@noop {} {\bibfield  {journal} {\bibinfo  {journal} {J.
  Chem. Phys.}\ }\textbf {\bibinfo {volume} {124}},\ \bibinfo {pages} {244704}
  (\bibinfo {year} {2006})}\BibitemShut {NoStop}%
\bibitem [{\citenamefont {Wang}\ \emph {et~al.}(2010)\citenamefont {Wang},
  \citenamefont {Lv}, \citenamefont {Zhu},\ and\ \citenamefont
  {Ma}}]{PhysRevB.82.094116}%
  \BibitemOpen
  \bibfield  {author} {\bibinfo {author} {\bibfnamefont {Y.}~\bibnamefont
  {Wang}}, \bibinfo {author} {\bibfnamefont {J.}~\bibnamefont {Lv}}, \bibinfo
  {author} {\bibfnamefont {L.}~\bibnamefont {Zhu}}, \ and\ \bibinfo {author}
  {\bibfnamefont {Y.}~\bibnamefont {Ma}},\ }\href@noop {} {\bibfield  {journal}
  {\bibinfo  {journal} {Phys. Rev. B}\ }\textbf {\bibinfo {volume} {82}},\
  \bibinfo {pages} {094116} (\bibinfo {year} {2010})}\BibitemShut {NoStop}%
\bibitem [{\citenamefont {Kim}\ \emph {et~al.}(2015)\citenamefont {Kim},
  \citenamefont {Stefanoski}, \citenamefont {Kurakevych},\ and\ \citenamefont
  {Strobel}}]{kim_synthesis_2015}%
  \BibitemOpen
  \bibfield  {author} {\bibinfo {author} {\bibfnamefont {D.~Y.}\ \bibnamefont
  {Kim}}, \bibinfo {author} {\bibfnamefont {S.}~\bibnamefont {Stefanoski}},
  \bibinfo {author} {\bibfnamefont {O.~O.}\ \bibnamefont {Kurakevych}}, \ and\
  \bibinfo {author} {\bibfnamefont {T.~A.}\ \bibnamefont {Strobel}},\ }\href
  {\doibase 10.1038/nmat4140} {\bibfield  {journal} {\bibinfo  {journal}
  {Nature Materials}\ }\textbf {\bibinfo {volume} {14}},\ \bibinfo {pages}
  {169} (\bibinfo {year} {2015})}\BibitemShut {NoStop}%
\bibitem [{\citenamefont {Davies}\ \emph {et~al.}(2018)\citenamefont {Davies},
  \citenamefont {Butler}, \citenamefont {Skelton}, \citenamefont {Xie},
  \citenamefont {Oganov},\ and\ \citenamefont
  {Walsh}}]{davies_computer-aided_2018}%
  \BibitemOpen
  \bibfield  {author} {\bibinfo {author} {\bibfnamefont {D.}~\bibnamefont
  {Davies}}, \bibinfo {author} {\bibfnamefont {K.~T.}\ \bibnamefont {Butler}},
  \bibinfo {author} {\bibfnamefont {J.~M.}\ \bibnamefont {Skelton}}, \bibinfo
  {author} {\bibfnamefont {C.}~\bibnamefont {Xie}}, \bibinfo {author}
  {\bibfnamefont {A.~R.}\ \bibnamefont {Oganov}}, \ and\ \bibinfo {author}
  {\bibfnamefont {A.}~\bibnamefont {Walsh}},\ }\href@noop {} {\bibfield
  {journal} {\bibinfo  {journal} {Chem. Sci.}\ }\textbf {\bibinfo {volume}
  {9}},\ \bibinfo {pages} {1022} (\bibinfo {year} {2018})}\BibitemShut
  {NoStop}%
\bibitem [{\citenamefont {Gu}\ \emph {et~al.}(2017)\citenamefont {Gu},
  \citenamefont {Luo},\ and\ \citenamefont {Xiang}}]{gu_prediction_2017}%
  \BibitemOpen
  \bibfield  {author} {\bibinfo {author} {\bibfnamefont {T.}~\bibnamefont
  {Gu}}, \bibinfo {author} {\bibfnamefont {W.}~\bibnamefont {Luo}}, \ and\
  \bibinfo {author} {\bibfnamefont {H.}~\bibnamefont {Xiang}},\ }\href
  {\doibase 10.1002/wcms.1295} {\bibfield  {journal} {\bibinfo  {journal}
  {WIREs Comput Mol Sci}\ }\textbf {\bibinfo {volume} {7}} (\bibinfo {year}
  {2017}),\ 10.1002/wcms.1295}\BibitemShut {NoStop}%
\bibitem [{\citenamefont {Kononova}\ \emph {et~al.}(2019)\citenamefont
  {Kononova}, \citenamefont {Huo}, \citenamefont {He}, \citenamefont {Rong},
  \citenamefont {Botari}, \citenamefont {Sun}, \citenamefont {Tshitoyan},\ and\
  \citenamefont {Ceder}}]{kononova_text-mined_2019}%
  \BibitemOpen
  \bibfield  {author} {\bibinfo {author} {\bibfnamefont {O.}~\bibnamefont
  {Kononova}}, \bibinfo {author} {\bibfnamefont {H.}~\bibnamefont {Huo}},
  \bibinfo {author} {\bibfnamefont {T.}~\bibnamefont {He}}, \bibinfo {author}
  {\bibfnamefont {Z.}~\bibnamefont {Rong}}, \bibinfo {author} {\bibfnamefont
  {T.}~\bibnamefont {Botari}}, \bibinfo {author} {\bibfnamefont
  {W.}~\bibnamefont {Sun}}, \bibinfo {author} {\bibfnamefont {V.}~\bibnamefont
  {Tshitoyan}}, \ and\ \bibinfo {author} {\bibfnamefont {G.}~\bibnamefont
  {Ceder}},\ }\href@noop {} {\bibfield  {journal} {\bibinfo  {journal} {Sci
  Data}\ }\textbf {\bibinfo {volume} {6}},\ \bibinfo {pages} {203} (\bibinfo
  {year} {2019})}\BibitemShut {NoStop}%
\bibitem [{\citenamefont {Yu}\ and\ \citenamefont
  {Zunger}(2012)}]{PhysRevLett_108_068701_2012}%
  \BibitemOpen
  \bibfield  {author} {\bibinfo {author} {\bibfnamefont {L.}~\bibnamefont
  {Yu}}\ and\ \bibinfo {author} {\bibfnamefont {A.}~\bibnamefont {Zunger}},\
  }\href@noop {} {\bibfield  {journal} {\bibinfo  {journal} {Phys. Rev. Lett.}\
  }\textbf {\bibinfo {volume} {108}},\ \bibinfo {pages} {068701} (\bibinfo
  {year} {2012})}\BibitemShut {NoStop}%
\bibitem [{\citenamefont {Choudhary}\ \emph {et~al.}(2019)\citenamefont
  {Choudhary}, \citenamefont {Bercx}, \citenamefont {Jiang}, \citenamefont
  {Pachter}, \citenamefont {Lamoen},\ and\ \citenamefont
  {Tavazza}}]{acs.chemmater.9b02166}%
  \BibitemOpen
  \bibfield  {author} {\bibinfo {author} {\bibfnamefont {K.}~\bibnamefont
  {Choudhary}}, \bibinfo {author} {\bibfnamefont {M.}~\bibnamefont {Bercx}},
  \bibinfo {author} {\bibfnamefont {J.}~\bibnamefont {Jiang}}, \bibinfo
  {author} {\bibfnamefont {R.}~\bibnamefont {Pachter}}, \bibinfo {author}
  {\bibfnamefont {D.}~\bibnamefont {Lamoen}}, \ and\ \bibinfo {author}
  {\bibfnamefont {F.}~\bibnamefont {Tavazza}},\ }\href@noop {} {\bibfield
  {journal} {\bibinfo  {journal} {Chemistry of Materials}\ }\textbf {\bibinfo
  {volume} {31}},\ \bibinfo {pages} {5900} (\bibinfo {year}
  {2019})}\BibitemShut {NoStop}%
\bibitem [{\citenamefont {Wang}\ \emph {et~al.}(2018)\citenamefont {Wang},
  \citenamefont {Li}, \citenamefont {Liang}, \citenamefont {Nie},\ and\
  \citenamefont {Wang}}]{wang_kagse_2018}%
  \BibitemOpen
  \bibfield  {author} {\bibinfo {author} {\bibfnamefont {Q.}~\bibnamefont
  {Wang}}, \bibinfo {author} {\bibfnamefont {J.}~\bibnamefont {Li}}, \bibinfo
  {author} {\bibfnamefont {Y.}~\bibnamefont {Liang}}, \bibinfo {author}
  {\bibfnamefont {Y.}~\bibnamefont {Nie}}, \ and\ \bibinfo {author}
  {\bibfnamefont {B.}~\bibnamefont {Wang}},\ }\href {\doibase
  10.1021/acsami.8b16505} {\bibfield  {journal} {\bibinfo  {journal} {ACS Appl.
  Mater. Interfaces}\ }\textbf {\bibinfo {volume} {10}},\ \bibinfo {pages}
  {41670} (\bibinfo {year} {2018})}\BibitemShut {NoStop}%
\bibitem [{\citenamefont {Yang}\ \emph {et~al.}(2019)\citenamefont {Yang},
  \citenamefont {Ma}, \citenamefont {Liang}, \citenamefont {Huang},\ and\
  \citenamefont {Dai}}]{yang_monolayer_2019}%
  \BibitemOpen
  \bibfield  {author} {\bibinfo {author} {\bibfnamefont {H.}~\bibnamefont
  {Yang}}, \bibinfo {author} {\bibfnamefont {Y.}~\bibnamefont {Ma}}, \bibinfo
  {author} {\bibfnamefont {Y.}~\bibnamefont {Liang}}, \bibinfo {author}
  {\bibfnamefont {B.}~\bibnamefont {Huang}}, \ and\ \bibinfo {author}
  {\bibfnamefont {Y.}~\bibnamefont {Dai}},\ }\href {\doibase
  10.1021/acsami.9b14920} {\bibfield  {journal} {\bibinfo  {journal} {ACS Appl.
  Mater. Interfaces}\ }\textbf {\bibinfo {volume} {11}},\ \bibinfo {pages}
  {37901} (\bibinfo {year} {2019})}\BibitemShut {NoStop}%
\bibitem [{\citenamefont {Zhao}\ \emph {et~al.}(2017)\citenamefont {Zhao},
  \citenamefont {Yang}, \citenamefont {Li}, \citenamefont {Jin}, \citenamefont
  {Wei}, \citenamefont {Yu}, \citenamefont {Huang},\ and\ \citenamefont
  {Dai}}]{zhao_design_2017}%
  \BibitemOpen
  \bibfield  {author} {\bibinfo {author} {\bibfnamefont {P.}~\bibnamefont
  {Zhao}}, \bibinfo {author} {\bibfnamefont {H.}~\bibnamefont {Yang}}, \bibinfo
  {author} {\bibfnamefont {J.}~\bibnamefont {Li}}, \bibinfo {author}
  {\bibfnamefont {H.}~\bibnamefont {Jin}}, \bibinfo {author} {\bibfnamefont
  {W.}~\bibnamefont {Wei}}, \bibinfo {author} {\bibfnamefont {L.}~\bibnamefont
  {Yu}}, \bibinfo {author} {\bibfnamefont {B.}~\bibnamefont {Huang}}, \ and\
  \bibinfo {author} {\bibfnamefont {Y.}~\bibnamefont {Dai}},\ }\href {\doibase
  10.1039/C7TA08097B} {\bibfield  {journal} {\bibinfo  {journal} {J. Mater.
  Chem. A}\ }\textbf {\bibinfo {volume} {5}},\ \bibinfo {pages} {24145}
  (\bibinfo {year} {2017})}\BibitemShut {NoStop}%
\bibitem [{\citenamefont {Chae}\ and\ \citenamefont
  {Son}(2019)}]{NanoLett_19_2694_2019}%
  \BibitemOpen
  \bibfield  {author} {\bibinfo {author} {\bibfnamefont {K.}~\bibnamefont
  {Chae}}\ and\ \bibinfo {author} {\bibfnamefont {Y.-W.}\ \bibnamefont {Son}},\
  }\href@noop {} {\bibfield  {journal} {\bibinfo  {journal} {Nano. Lett.}\
  }\textbf {\bibinfo {volume} {19}},\ \bibinfo {pages} {2694} (\bibinfo {year}
  {2019})}\BibitemShut {NoStop}%
\bibitem [{\citenamefont {Chae}\ \emph {et~al.}(2018)\citenamefont {Chae},
  \citenamefont {Kim},\ and\ \citenamefont {Son}}]{chae_new_2018}%
  \BibitemOpen
  \bibfield  {author} {\bibinfo {author} {\bibfnamefont {K.}~\bibnamefont
  {Chae}}, \bibinfo {author} {\bibfnamefont {D.~Y.}\ \bibnamefont {Kim}}, \
  and\ \bibinfo {author} {\bibfnamefont {Y.-W.}\ \bibnamefont {Son}},\ }\href
  {\doibase 10.1088/2053-1583/aaac9f} {\bibfield  {journal} {\bibinfo
  {journal} {2D Mater.}\ }\textbf {\bibinfo {volume} {5}},\ \bibinfo {pages}
  {025013} (\bibinfo {year} {2018})}\BibitemShut {NoStop}%
\bibitem [{\citenamefont {Mak}\ \emph {et~al.}(2010)\citenamefont {Mak},
  \citenamefont {Lee}, \citenamefont {Hone}, \citenamefont {Shan},\ and\
  \citenamefont {Heinz}}]{PhysRevLett.105.136805}%
  \BibitemOpen
  \bibfield  {author} {\bibinfo {author} {\bibfnamefont {K.~F.}\ \bibnamefont
  {Mak}}, \bibinfo {author} {\bibfnamefont {C.}~\bibnamefont {Lee}}, \bibinfo
  {author} {\bibfnamefont {J.}~\bibnamefont {Hone}}, \bibinfo {author}
  {\bibfnamefont {J.}~\bibnamefont {Shan}}, \ and\ \bibinfo {author}
  {\bibfnamefont {T.~F.}\ \bibnamefont {Heinz}},\ }\href {\doibase
  10.1103/PhysRevLett.105.136805} {\bibfield  {journal} {\bibinfo  {journal}
  {Phys. Rev. Lett.}\ }\textbf {\bibinfo {volume} {105}},\ \bibinfo {pages}
  {136805} (\bibinfo {year} {2010})}\BibitemShut {NoStop}%
\bibitem [{\citenamefont {Momma}\ and\ \citenamefont
  {Izumi}(2006)}]{IUCr_Newslett_7106-119_2006}%
  \BibitemOpen
  \bibfield  {author} {\bibinfo {author} {\bibfnamefont {K.}~\bibnamefont
  {Momma}}\ and\ \bibinfo {author} {\bibfnamefont {F.}~\bibnamefont {Izumi}},\
  }\href@noop {} {\bibfield  {journal} {\bibinfo  {journal} {Commission on
  Crystallogr. Comput., IUCr Newslett.}\ }\textbf {\bibinfo {volume} {7}},\
  \bibinfo {pages} {106} (\bibinfo {year} {2006})}\BibitemShut {NoStop}%
\bibitem [{\citenamefont {Perdew}\ \emph {et~al.}(1996)\citenamefont {Perdew},
  \citenamefont {Burke},\ and\ \citenamefont
  {Ernzerhof}}]{PhysRevLett_77_3865_1996}%
  \BibitemOpen
  \bibfield  {author} {\bibinfo {author} {\bibfnamefont {J.~P.}\ \bibnamefont
  {Perdew}}, \bibinfo {author} {\bibfnamefont {K.}~\bibnamefont {Burke}}, \
  and\ \bibinfo {author} {\bibfnamefont {M.}~\bibnamefont {Ernzerhof}},\
  }\href@noop {} {\bibfield  {journal} {\bibinfo  {journal} {Phys. Rev. Lett.}\
  }\textbf {\bibinfo {volume} {77}},\ \bibinfo {pages} {3865} (\bibinfo {year}
  {1996})}\BibitemShut {NoStop}%
\bibitem [{\citenamefont {Giannozzi}\ \emph {et~al.}(2009)\citenamefont
  {Giannozzi}, \citenamefont {Andreussi}, \citenamefont {Brumme}, \citenamefont
  {Bunau}, \citenamefont {Nardelli}, \citenamefont {Calandra}, \citenamefont
  {Car}, \citenamefont {Cavazzoni}, \citenamefont {Ceresoli}, \citenamefont
  {Cococcioni}, \citenamefont {Colonna}, \citenamefont {Carnimeo},
  \citenamefont {Corso}, \citenamefont {de~Gironcoli}, \citenamefont {Delugas},
  \citenamefont {Jr}, \citenamefont {Ferretti}, \citenamefont {Floris},
  \citenamefont {Fratesi}, \citenamefont {Fugallo}, \citenamefont {Gebauer},
  \citenamefont {Gerstmann}, \citenamefont {Giustino}, \citenamefont {Gorni},
  \citenamefont {Jia}, \citenamefont {Kawamura}, \citenamefont {Ko},
  \citenamefont {Kokalj}, \citenamefont {K\"{u}c\"{u}kbenli}, \citenamefont
  {.Lazzeri}, \citenamefont {Marsili}, \citenamefont {Marzari}, \citenamefont
  {Mauri}, \citenamefont {Nguyen}, \citenamefont {Nguyen}, \citenamefont {de-la
  Roza}, \citenamefont {Paulatto}, \citenamefont {Ponc\'{e}}, \citenamefont
  {Rocca}, \citenamefont {Sabatini}, \citenamefont {Santra}, \citenamefont
  {Schlipf}, \citenamefont {Seitsonen}, \citenamefont {Smogunov}, \citenamefont
  {Timrov}, \citenamefont {Thonhauser}, \citenamefont {Umari}, \citenamefont
  {Vast}, \citenamefont {Wu},\ and\ \citenamefont
  {Baroni}}]{JPhys_CM_21_395502_2009}%
  \BibitemOpen
  \bibfield  {author} {\bibinfo {author} {\bibfnamefont {P.}~\bibnamefont
  {Giannozzi}}, \bibinfo {author} {\bibfnamefont {O.}~\bibnamefont
  {Andreussi}}, \bibinfo {author} {\bibfnamefont {T.}~\bibnamefont {Brumme}},
  \bibinfo {author} {\bibfnamefont {O.}~\bibnamefont {Bunau}}, \bibinfo
  {author} {\bibfnamefont {M.~B.}\ \bibnamefont {Nardelli}}, \bibinfo {author}
  {\bibfnamefont {M.}~\bibnamefont {Calandra}}, \bibinfo {author}
  {\bibfnamefont {R.}~\bibnamefont {Car}}, \bibinfo {author} {\bibfnamefont
  {C.}~\bibnamefont {Cavazzoni}}, \bibinfo {author} {\bibfnamefont
  {D.}~\bibnamefont {Ceresoli}}, \bibinfo {author} {\bibfnamefont
  {M.}~\bibnamefont {Cococcioni}}, \bibinfo {author} {\bibfnamefont
  {N.}~\bibnamefont {Colonna}}, \bibinfo {author} {\bibfnamefont
  {I.}~\bibnamefont {Carnimeo}}, \bibinfo {author} {\bibfnamefont {A.~D.}\
  \bibnamefont {Corso}}, \bibinfo {author} {\bibfnamefont {S.}~\bibnamefont
  {de~Gironcoli}}, \bibinfo {author} {\bibfnamefont {P.}~\bibnamefont
  {Delugas}}, \bibinfo {author} {\bibfnamefont {R.~A.~D.}\ \bibnamefont {Jr}},
  \bibinfo {author} {\bibfnamefont {A.}~\bibnamefont {Ferretti}}, \bibinfo
  {author} {\bibfnamefont {A.}~\bibnamefont {Floris}}, \bibinfo {author}
  {\bibfnamefont {G.}~\bibnamefont {Fratesi}}, \bibinfo {author} {\bibfnamefont
  {G.}~\bibnamefont {Fugallo}}, \bibinfo {author} {\bibfnamefont
  {R.}~\bibnamefont {Gebauer}}, \bibinfo {author} {\bibfnamefont
  {U.}~\bibnamefont {Gerstmann}}, \bibinfo {author} {\bibfnamefont
  {F.}~\bibnamefont {Giustino}}, \bibinfo {author} {\bibfnamefont
  {T.}~\bibnamefont {Gorni}}, \bibinfo {author} {\bibfnamefont
  {J.}~\bibnamefont {Jia}}, \bibinfo {author} {\bibfnamefont {M.}~\bibnamefont
  {Kawamura}}, \bibinfo {author} {\bibfnamefont {H.-Y.}\ \bibnamefont {Ko}},
  \bibinfo {author} {\bibfnamefont {A.}~\bibnamefont {Kokalj}}, \bibinfo
  {author} {\bibfnamefont {E.}~\bibnamefont {K\"{u}c\"{u}kbenli}}, \bibinfo
  {author} {\bibfnamefont {M.}~\bibnamefont {.Lazzeri}}, \bibinfo {author}
  {\bibfnamefont {M.}~\bibnamefont {Marsili}}, \bibinfo {author} {\bibfnamefont
  {N.}~\bibnamefont {Marzari}}, \bibinfo {author} {\bibfnamefont
  {F.}~\bibnamefont {Mauri}}, \bibinfo {author} {\bibfnamefont {N.~L.}\
  \bibnamefont {Nguyen}}, \bibinfo {author} {\bibfnamefont {H.-V.}\
  \bibnamefont {Nguyen}}, \bibinfo {author} {\bibfnamefont {A.~O.}\
  \bibnamefont {de-la Roza}}, \bibinfo {author} {\bibfnamefont
  {L.}~\bibnamefont {Paulatto}}, \bibinfo {author} {\bibfnamefont
  {S.}~\bibnamefont {Ponc\'{e}}}, \bibinfo {author} {\bibfnamefont
  {D.}~\bibnamefont {Rocca}}, \bibinfo {author} {\bibfnamefont
  {R.}~\bibnamefont {Sabatini}}, \bibinfo {author} {\bibfnamefont
  {B.}~\bibnamefont {Santra}}, \bibinfo {author} {\bibfnamefont
  {M.}~\bibnamefont {Schlipf}}, \bibinfo {author} {\bibfnamefont {A.~P.}\
  \bibnamefont {Seitsonen}}, \bibinfo {author} {\bibfnamefont {A.}~\bibnamefont
  {Smogunov}}, \bibinfo {author} {\bibfnamefont {I.}~\bibnamefont {Timrov}},
  \bibinfo {author} {\bibfnamefont {T.}~\bibnamefont {Thonhauser}}, \bibinfo
  {author} {\bibfnamefont {P.}~\bibnamefont {Umari}}, \bibinfo {author}
  {\bibfnamefont {N.}~\bibnamefont {Vast}}, \bibinfo {author} {\bibfnamefont
  {X.}~\bibnamefont {Wu}}, \ and\ \bibinfo {author} {\bibfnamefont
  {S.}~\bibnamefont {Baroni}},\ }\href@noop {} {\bibfield  {journal} {\bibinfo
  {journal} {J. Phys.: Condens. Matter}\ }\textbf {\bibinfo {volume} {21}},\
  \bibinfo {pages} {395502} (\bibinfo {year} {2009})}\BibitemShut {NoStop}%
\bibitem [{\citenamefont {Giannozzi}\ \emph {et~al.}(2017)\citenamefont
  {Giannozzi}, \citenamefont {Baroni}, \citenamefont {Bonini}, \citenamefont
  {Calandra}, \citenamefont {Car}, \citenamefont {Cavazzoni}, \citenamefont
  {Ceresoli}, \citenamefont {Chiarotti}, \citenamefont {Cococcioni},
  \citenamefont {amd A.~Dal~Corso}, \citenamefont {Fabris}, \citenamefont
  {Fratesi}, \citenamefont {de~Gironcoli}, \citenamefont {Gebauer},
  \citenamefont {Gerstmann}, \citenamefont {Gougoussis}, \citenamefont
  {Kokalj}, \citenamefont {Lazzeri}, \citenamefont {Martin-Samos},
  \citenamefont {Marzari}, \citenamefont {Mauri}, \citenamefont {Mazzarello},
  \citenamefont {Paolini}, \citenamefont {Pasquarello}, \citenamefont
  {Paulatto}, \citenamefont {Sbraccia}, \citenamefont {Scandolo}, \citenamefont
  {Sclauzero}, \citenamefont {Seitsonen}, \citenamefont {Smogunov},
  \citenamefont {Umari},\ and\ \citenamefont
  {Wentzcovitch}}]{JPhys_CM_29_465901_2017}%
  \BibitemOpen
  \bibfield  {author} {\bibinfo {author} {\bibfnamefont {P.}~\bibnamefont
  {Giannozzi}}, \bibinfo {author} {\bibfnamefont {S.}~\bibnamefont {Baroni}},
  \bibinfo {author} {\bibfnamefont {N.}~\bibnamefont {Bonini}}, \bibinfo
  {author} {\bibfnamefont {M.}~\bibnamefont {Calandra}}, \bibinfo {author}
  {\bibfnamefont {R.}~\bibnamefont {Car}}, \bibinfo {author} {\bibfnamefont
  {C.}~\bibnamefont {Cavazzoni}}, \bibinfo {author} {\bibfnamefont
  {D.}~\bibnamefont {Ceresoli}}, \bibinfo {author} {\bibfnamefont {G.~L.}\
  \bibnamefont {Chiarotti}}, \bibinfo {author} {\bibfnamefont {M.}~\bibnamefont
  {Cococcioni}}, \bibinfo {author} {\bibfnamefont {I.~D.}\ \bibnamefont {amd
  A.~Dal~Corso}}, \bibinfo {author} {\bibfnamefont {S.}~\bibnamefont {Fabris}},
  \bibinfo {author} {\bibfnamefont {G.}~\bibnamefont {Fratesi}}, \bibinfo
  {author} {\bibfnamefont {S.}~\bibnamefont {de~Gironcoli}}, \bibinfo {author}
  {\bibfnamefont {R.}~\bibnamefont {Gebauer}}, \bibinfo {author} {\bibfnamefont
  {U.}~\bibnamefont {Gerstmann}}, \bibinfo {author} {\bibfnamefont
  {C.}~\bibnamefont {Gougoussis}}, \bibinfo {author} {\bibfnamefont
  {A.}~\bibnamefont {Kokalj}}, \bibinfo {author} {\bibfnamefont
  {M.}~\bibnamefont {Lazzeri}}, \bibinfo {author} {\bibfnamefont
  {L.}~\bibnamefont {Martin-Samos}}, \bibinfo {author} {\bibfnamefont
  {N.}~\bibnamefont {Marzari}}, \bibinfo {author} {\bibfnamefont
  {F.}~\bibnamefont {Mauri}}, \bibinfo {author} {\bibfnamefont
  {R.}~\bibnamefont {Mazzarello}}, \bibinfo {author} {\bibfnamefont
  {S.}~\bibnamefont {Paolini}}, \bibinfo {author} {\bibfnamefont
  {A.}~\bibnamefont {Pasquarello}}, \bibinfo {author} {\bibfnamefont
  {L.}~\bibnamefont {Paulatto}}, \bibinfo {author} {\bibfnamefont
  {C.}~\bibnamefont {Sbraccia}}, \bibinfo {author} {\bibfnamefont
  {S.}~\bibnamefont {Scandolo}}, \bibinfo {author} {\bibfnamefont
  {G.}~\bibnamefont {Sclauzero}}, \bibinfo {author} {\bibfnamefont {A.~P.}\
  \bibnamefont {Seitsonen}}, \bibinfo {author} {\bibfnamefont {A.}~\bibnamefont
  {Smogunov}}, \bibinfo {author} {\bibfnamefont {P.}~\bibnamefont {Umari}}, \
  and\ \bibinfo {author} {\bibfnamefont {R.~M.}\ \bibnamefont {Wentzcovitch}},\
  }\href@noop {} {\bibfield  {journal} {\bibinfo  {journal} {J. Phys.: Condens.
  Matter}\ }\textbf {\bibinfo {volume} {29}},\ \bibinfo {pages} {465901}
  (\bibinfo {year} {2017})}\BibitemShut {NoStop}%
\bibitem [{\citenamefont {Grimme}(2006)}]{J_Comp_Chem_27_1787_2006}%
  \BibitemOpen
  \bibfield  {author} {\bibinfo {author} {\bibfnamefont {S.}~\bibnamefont
  {Grimme}},\ }\href@noop {} {\bibfield  {journal} {\bibinfo  {journal} {J.
  Comp. Chem.}\ }\textbf {\bibinfo {volume} {27}},\ \bibinfo {pages} {1787}
  (\bibinfo {year} {2006})}\BibitemShut {NoStop}%
\bibitem [{\citenamefont {Hedin}(1965)}]{PhysRev_139_A796_1965}%
  \BibitemOpen
  \bibfield  {author} {\bibinfo {author} {\bibfnamefont {L.}~\bibnamefont
  {Hedin}},\ }\href@noop {} {\bibfield  {journal} {\bibinfo  {journal} {Phys.
  Rev.}\ }\textbf {\bibinfo {volume} {139}},\ \bibinfo {pages} {A796} (\bibinfo
  {year} {1965})}\BibitemShut {NoStop}%
\bibitem [{\citenamefont {Deslippe}\ \emph {et~al.}(2013)\citenamefont
  {Deslippe}, \citenamefont {Samsonidze}, \citenamefont {Jain}, \citenamefont
  {Cohen},\ and\ \citenamefont {Louie}}]{PhysRevB_87_165124_2013}%
  \BibitemOpen
  \bibfield  {author} {\bibinfo {author} {\bibfnamefont {J.}~\bibnamefont
  {Deslippe}}, \bibinfo {author} {\bibfnamefont {G.}~\bibnamefont
  {Samsonidze}}, \bibinfo {author} {\bibfnamefont {M.}~\bibnamefont {Jain}},
  \bibinfo {author} {\bibfnamefont {M.~L.}\ \bibnamefont {Cohen}}, \ and\
  \bibinfo {author} {\bibfnamefont {S.~G.}\ \bibnamefont {Louie}},\ }\href@noop
  {} {\bibfield  {journal} {\bibinfo  {journal} {Phys. Rev. B}\ }\textbf
  {\bibinfo {volume} {87}},\ \bibinfo {pages} {165124} (\bibinfo {year}
  {2013})}\BibitemShut {NoStop}%
\bibitem [{\citenamefont {Salpeter}\ and\ \citenamefont
  {Bethe}(1951)}]{PhysRev_84_1232_1951}%
  \BibitemOpen
  \bibfield  {author} {\bibinfo {author} {\bibfnamefont {E.~E.}\ \bibnamefont
  {Salpeter}}\ and\ \bibinfo {author} {\bibfnamefont {H.~A.}\ \bibnamefont
  {Bethe}},\ }\href@noop {} {\bibfield  {journal} {\bibinfo  {journal} {Phys.
  Rev.}\ }\textbf {\bibinfo {volume} {84}},\ \bibinfo {pages} {1232} (\bibinfo
  {year} {1951})}\BibitemShut {NoStop}%
\bibitem [{\citenamefont {Albrecht}\ \emph {et~al.}(1998)\citenamefont
  {Albrecht}, \citenamefont {Reining}, \citenamefont {Sole},\ and\
  \citenamefont {Onida}}]{PhysRevLett_80_4510_1998}%
  \BibitemOpen
  \bibfield  {author} {\bibinfo {author} {\bibfnamefont {S.}~\bibnamefont
  {Albrecht}}, \bibinfo {author} {\bibfnamefont {L.}~\bibnamefont {Reining}},
  \bibinfo {author} {\bibfnamefont {R.~D.}\ \bibnamefont {Sole}}, \ and\
  \bibinfo {author} {\bibfnamefont {G.}~\bibnamefont {Onida}},\ }\href@noop {}
  {\bibfield  {journal} {\bibinfo  {journal} {Phys. Rev. Lett.}\ }\textbf
  {\bibinfo {volume} {80}},\ \bibinfo {pages} {4510} (\bibinfo {year}
  {1998})}\BibitemShut {NoStop}%
\bibitem [{\citenamefont {Benedict}\ \emph {et~al.}(1998)\citenamefont
  {Benedict}, \citenamefont {Shirley},\ and\ \citenamefont
  {Bohn}}]{PhysRevLett_80_4514_1998}%
  \BibitemOpen
  \bibfield  {author} {\bibinfo {author} {\bibfnamefont {L.~X.}\ \bibnamefont
  {Benedict}}, \bibinfo {author} {\bibfnamefont {E.~L.}\ \bibnamefont
  {Shirley}}, \ and\ \bibinfo {author} {\bibfnamefont {R.~B.}\ \bibnamefont
  {Bohn}},\ }\href@noop {} {\bibfield  {journal} {\bibinfo  {journal} {Phys.
  Rev. Lett.}\ }\textbf {\bibinfo {volume} {80}},\ \bibinfo {pages} {4514}
  (\bibinfo {year} {1998})}\BibitemShut {NoStop}%
\bibitem [{\citenamefont {Hybertsen}\ and\ \citenamefont
  {Louie}(1986)}]{PhysRevB_34_5390_1986}%
  \BibitemOpen
  \bibfield  {author} {\bibinfo {author} {\bibfnamefont {M.}~\bibnamefont
  {Hybertsen}}\ and\ \bibinfo {author} {\bibfnamefont {S.}~\bibnamefont
  {Louie}},\ }\href@noop {} {\bibfield  {journal} {\bibinfo  {journal} {Phys.
  Rev. B}\ }\textbf {\bibinfo {volume} {34}},\ \bibinfo {pages} {5390}
  (\bibinfo {year} {1986})}\BibitemShut {NoStop}%
\bibitem [{\citenamefont {Rohlfing}\ and\ \citenamefont
  {Louie}(1998)}]{PhysRevLett_81_2312_1998}%
  \BibitemOpen
  \bibfield  {author} {\bibinfo {author} {\bibfnamefont {M.}~\bibnamefont
  {Rohlfing}}\ and\ \bibinfo {author} {\bibfnamefont {S.~G.}\ \bibnamefont
  {Louie}},\ }\href@noop {} {\bibfield  {journal} {\bibinfo  {journal} {Phys.
  Rev. Lett.}\ }\textbf {\bibinfo {volume} {81}},\ \bibinfo {pages} {2312}
  (\bibinfo {year} {1998})}\BibitemShut {NoStop}%
\bibitem [{\citenamefont {Rohlfing}\ and\ \citenamefont
  {Louie}(2000)}]{PhysRevB_62_4927_2000}%
  \BibitemOpen
  \bibfield  {author} {\bibinfo {author} {\bibfnamefont {M.}~\bibnamefont
  {Rohlfing}}\ and\ \bibinfo {author} {\bibfnamefont {S.~G.}\ \bibnamefont
  {Louie}},\ }\href@noop {} {\bibfield  {journal} {\bibinfo  {journal} {Phys.
  Rev. B}\ }\textbf {\bibinfo {volume} {62}},\ \bibinfo {pages} {4927}
  (\bibinfo {year} {2000})}\BibitemShut {NoStop}%
\bibitem [{\citenamefont {Deslippe}\ \emph {et~al.}(2012)\citenamefont
  {Deslippe}, \citenamefont {Samsonidze}, \citenamefont {Strubbe},
  \citenamefont {Jain}, \citenamefont {Cohen},\ and\ \citenamefont
  {Louie}}]{Com_Phys_Comm_183_1269_2012}%
  \BibitemOpen
  \bibfield  {author} {\bibinfo {author} {\bibfnamefont {J.}~\bibnamefont
  {Deslippe}}, \bibinfo {author} {\bibfnamefont {G.}~\bibnamefont
  {Samsonidze}}, \bibinfo {author} {\bibfnamefont {D.~A.}\ \bibnamefont
  {Strubbe}}, \bibinfo {author} {\bibfnamefont {M.}~\bibnamefont {Jain}},
  \bibinfo {author} {\bibfnamefont {M.~L.}\ \bibnamefont {Cohen}}, \ and\
  \bibinfo {author} {\bibfnamefont {S.~G.}\ \bibnamefont {Louie}},\ }\href@noop
  {} {\bibfield  {journal} {\bibinfo  {journal} {Comput. Phys. Commun.}\
  }\textbf {\bibinfo {volume} {183}},\ \bibinfo {pages} {1269} (\bibinfo {year}
  {2012})}\BibitemShut {NoStop}%
\bibitem [{\citenamefont {Ismail-Beigi}(2006)}]{PhysRevB_73_233103_2006}%
  \BibitemOpen
  \bibfield  {author} {\bibinfo {author} {\bibfnamefont {S.}~\bibnamefont
  {Ismail-Beigi}},\ }\href@noop {} {\bibfield  {journal} {\bibinfo  {journal}
  {Phys. Rev. B}\ }\textbf {\bibinfo {volume} {73}},\ \bibinfo {pages} {233103}
  (\bibinfo {year} {2006})}\BibitemShut {NoStop}%
\bibitem [{\citenamefont {Qiu}\ \emph {et~al.}(2013)\citenamefont {Qiu},
  \citenamefont {da~Jornada},\ and\ \citenamefont
  {Louie}}]{PhysRevLett_111_216805_2013}%
  \BibitemOpen
  \bibfield  {author} {\bibinfo {author} {\bibfnamefont {D.~Y.}\ \bibnamefont
  {Qiu}}, \bibinfo {author} {\bibfnamefont {F.~H.}\ \bibnamefont {da~Jornada}},
  \ and\ \bibinfo {author} {\bibfnamefont {S.~G.}\ \bibnamefont {Louie}},\
  }\href@noop {} {\bibfield  {journal} {\bibinfo  {journal} {Phys. Rev. Lett.}\
  }\textbf {\bibinfo {volume} {111}},\ \bibinfo {pages} {216805} (\bibinfo
  {year} {2013})}\BibitemShut {NoStop}%
\bibitem [{\citenamefont {Heyd}\ \emph {et~al.}(2003)\citenamefont {Heyd},
  \citenamefont {Scuseria},\ and\ \citenamefont
  {Ernzerhof}}]{JChemPhys_118_8207_2003}%
  \BibitemOpen
  \bibfield  {author} {\bibinfo {author} {\bibfnamefont {J.}~\bibnamefont
  {Heyd}}, \bibinfo {author} {\bibfnamefont {G.~E.}\ \bibnamefont {Scuseria}},
  \ and\ \bibinfo {author} {\bibfnamefont {M.}~\bibnamefont {Ernzerhof}},\
  }\href@noop {} {\bibfield  {journal} {\bibinfo  {journal} {J. Chem. Phys.}\
  }\textbf {\bibinfo {volume} {118}},\ \bibinfo {pages} {8207} (\bibinfo {year}
  {2003})}\BibitemShut {NoStop}%
\bibitem [{\citenamefont {Krukau}\ \emph {et~al.}(2006)\citenamefont {Krukau},
  \citenamefont {Vydrov}, \citenamefont {Izmaylov},\ and\ \citenamefont
  {Scuseria}}]{JChemPhys_125_224106_2006}%
  \BibitemOpen
  \bibfield  {author} {\bibinfo {author} {\bibfnamefont {A.~V.}\ \bibnamefont
  {Krukau}}, \bibinfo {author} {\bibfnamefont {O.~A.}\ \bibnamefont {Vydrov}},
  \bibinfo {author} {\bibfnamefont {A.~F.}\ \bibnamefont {Izmaylov}}, \ and\
  \bibinfo {author} {\bibfnamefont {G.~E.}\ \bibnamefont {Scuseria}},\
  }\href@noop {} {\bibfield  {journal} {\bibinfo  {journal} {J. Chem. Phys.}\
  }\textbf {\bibinfo {volume} {125}},\ \bibinfo {pages} {224106} (\bibinfo
  {year} {2006})}\BibitemShut {NoStop}%
\bibitem [{\citenamefont {Qiu}\ \emph {et~al.}(2016)\citenamefont {Qiu},
  \citenamefont {da~Jornada},\ and\ \citenamefont
  {Louie}}]{PhysRevB.93.235435}%
  \BibitemOpen
  \bibfield  {author} {\bibinfo {author} {\bibfnamefont {D.~Y.}\ \bibnamefont
  {Qiu}}, \bibinfo {author} {\bibfnamefont {F.~H.}\ \bibnamefont {da~Jornada}},
  \ and\ \bibinfo {author} {\bibfnamefont {S.~G.}\ \bibnamefont {Louie}},\
  }\href@noop {} {\bibfield  {journal} {\bibinfo  {journal} {Phys. Rev. B}\
  }\textbf {\bibinfo {volume} {93}},\ \bibinfo {pages} {235435} (\bibinfo
  {year} {2016})}\BibitemShut {NoStop}%
\bibitem [{\citenamefont {Shockley}\ and\ \citenamefont
  {Queisser}(1961)}]{JApplPhys_32_510_1961}%
  \BibitemOpen
  \bibfield  {author} {\bibinfo {author} {\bibfnamefont {W.}~\bibnamefont
  {Shockley}}\ and\ \bibinfo {author} {\bibfnamefont {H.~J.}\ \bibnamefont
  {Queisser}},\ }\href@noop {} {\bibfield  {journal} {\bibinfo  {journal} {J.
  Appl. Phys.}\ }\textbf {\bibinfo {volume} {32}},\ \bibinfo {pages} {510}
  (\bibinfo {year} {1961})}\BibitemShut {NoStop}%
\bibitem [{\citenamefont {Marder}(2010)}]{Condensed_Matter_Physics_Marder}%
  \BibitemOpen
  \bibfield  {author} {\bibinfo {author} {\bibfnamefont {M.~P.}\ \bibnamefont
  {Marder}},\ }\href@noop {} {\emph {\bibinfo {title} {Condensed Matter
  Physics}}},\ \bibinfo {edition} {2nd}\ ed.\ (\bibinfo  {publisher} {John
  Wiley \& Sons, Inc.},\ \bibinfo {year} {2010})\BibitemShut {NoStop}%
\bibitem [{\citenamefont {Bernardi}\ \emph {et~al.}(2013)\citenamefont
  {Bernardi}, \citenamefont {Palummo},\ and\ \citenamefont
  {Grossman}}]{bernardi_extraordinary_2013}%
  \BibitemOpen
  \bibfield  {author} {\bibinfo {author} {\bibfnamefont {M.}~\bibnamefont
  {Bernardi}}, \bibinfo {author} {\bibfnamefont {M.}~\bibnamefont {Palummo}}, \
  and\ \bibinfo {author} {\bibfnamefont {J.~C.}\ \bibnamefont {Grossman}},\
  }\href {\doibase 10.1021/nl401544y} {\bibfield  {journal} {\bibinfo
  {journal} {Nano Lett.}\ }\textbf {\bibinfo {volume} {13}},\ \bibinfo {pages}
  {3664} (\bibinfo {year} {2013})}\BibitemShut {NoStop}%
\bibitem [{AM1()}]{AM1.5}%
  \BibitemOpen
  \href@noop {} {\enquote {\bibinfo {title} {Reference solar spectral
  irradiance: Air mass 1.5},}\ }\bibinfo {howpublished}
  {\url{https://rredc.nrel.gov/solar//spectra/am1.5/}}\BibitemShut {NoStop}%
\end{thebibliography}

\begin{thebibliography}{16}%
\makeatletter
\providecommand \@ifxundefined [1]{%
 \@ifx{#1\undefined}
}%
\providecommand \@ifnum [1]{%
 \ifnum #1\expandafter \@firstoftwo
 \else \expandafter \@secondoftwo
 \fi
}%
\providecommand \@ifx [1]{%
 \ifx #1\expandafter \@firstoftwo
 \else \expandafter \@secondoftwo
 \fi
}%
\providecommand \natexlab [1]{#1}%
\providecommand \enquote  [1]{``#1''}%
\providecommand \bibnamefont  [1]{#1}%
\providecommand \bibfnamefont [1]{#1}%
\providecommand \citenamefont [1]{#1}%
\providecommand \href@noop [0]{\@secondoftwo}%
\providecommand \href [0]{\begingroup \@sanitize@url \@href}%
\providecommand \@href[1]{\@@startlink{#1}\@@href}%
\providecommand \@@href[1]{\endgroup#1\@@endlink}%
\providecommand \@sanitize@url [0]{\catcode `\\12\catcode `\$12\catcode
  `\&12\catcode `\#12\catcode `\^12\catcode `\_12\catcode `\%12\relax}%
\providecommand \@@startlink[1]{}%
\providecommand \@@endlink[0]{}%
\providecommand \url  [0]{\begingroup\@sanitize@url \@url }%
\providecommand \@url [1]{\endgroup\@href {#1}{\urlprefix }}%
\providecommand \urlprefix  [0]{URL }%
\providecommand \Eprint [0]{\href }%
\providecommand \doibase [0]{http://dx.doi.org/}%
\providecommand \selectlanguage [0]{\@gobble}%
\providecommand \bibinfo  [0]{\@secondoftwo}%
\providecommand \bibfield  [0]{\@secondoftwo}%
\providecommand \translation [1]{[#1]}%
\providecommand \BibitemOpen [0]{}%
\providecommand \bibitemStop [0]{}%
\providecommand \bibitemNoStop [0]{.\EOS\space}%
\providecommand \EOS [0]{\spacefactor3000\relax}%
\providecommand \BibitemShut  [1]{\csname bibitem#1\endcsname}%
\let\auto@bib@innerbib\@empty
%</preamble>
\bibitem [{\citenamefont {Giannozzi}\ \emph {et~al.}(2009)\citenamefont
  {Giannozzi}, \citenamefont {Andreussi}, \citenamefont {Brumme}, \citenamefont
  {Bunau}, \citenamefont {Nardelli}, \citenamefont {Calandra}, \citenamefont
  {Car}, \citenamefont {Cavazzoni}, \citenamefont {Ceresoli}, \citenamefont
  {Cococcioni}, \citenamefont {Colonna}, \citenamefont {Carnimeo},
  \citenamefont {Corso}, \citenamefont {de~Gironcoli}, \citenamefont {Delugas},
  \citenamefont {Jr}, \citenamefont {Ferretti}, \citenamefont {Floris},
  \citenamefont {Fratesi}, \citenamefont {Fugallo}, \citenamefont {Gebauer},
  \citenamefont {Gerstmann}, \citenamefont {Giustino}, \citenamefont {Gorni},
  \citenamefont {Jia}, \citenamefont {Kawamura}, \citenamefont {Ko},
  \citenamefont {Kokalj}, \citenamefont {K\"{u}c\"{u}kbenli}, \citenamefont
  {Lazzeri}, \citenamefont {Marsili}, \citenamefont {Marzari}, \citenamefont
  {Mauri}, \citenamefont {Nguyen}, \citenamefont {Nguyen}, \citenamefont {de-la
  Roza}, \citenamefont {Paulatto}, \citenamefont {Ponc\'{e}}, \citenamefont
  {Rocca}, \citenamefont {Sabatini}, \citenamefont {Santra}, \citenamefont
  {Schlipf}, \citenamefont {Seitsonen}, \citenamefont {Smogunov}, \citenamefont
  {Timrov}, \citenamefont {Thonhauser}, \citenamefont {Umari}, \citenamefont
  {Vast}, \citenamefont {Wu},\ and\ \citenamefont
  {Baroni}}]{JPhys_CM_21_395502_2009-2}%
  \BibitemOpen
  \bibfield  {author} {\bibinfo {author} {\bibfnamefont {P.}~\bibnamefont
  {Giannozzi}}, \bibinfo {author} {\bibfnamefont {O.}~\bibnamefont
  {Andreussi}}, \bibinfo {author} {\bibfnamefont {T.}~\bibnamefont {Brumme}},
  \bibinfo {author} {\bibfnamefont {O.}~\bibnamefont {Bunau}}, \bibinfo
  {author} {\bibfnamefont {M.~Buongiorno}\ \bibnamefont {Nardelli}}, \bibinfo
  {author} {\bibfnamefont {M.}~\bibnamefont {Calandra}}, \bibinfo {author}
  {\bibfnamefont {R.}~\bibnamefont {Car}}, \bibinfo {author} {\bibfnamefont
  {C.}~\bibnamefont {Cavazzoni}}, \bibinfo {author} {\bibfnamefont
  {D.}~\bibnamefont {Ceresoli}}, \bibinfo {author} {\bibfnamefont
  {M.}~\bibnamefont {Cococcioni}}, \bibinfo {author} {\bibfnamefont
  {N.}~\bibnamefont {Colonna}}, \bibinfo {author} {\bibfnamefont
  {I.}~\bibnamefont {Carnimeo}}, \bibinfo {author} {\bibfnamefont {A.~Dal}\
  \bibnamefont {Corso}}, \bibinfo {author} {\bibfnamefont {S.}~\bibnamefont
  {de~Gironcoli}}, \bibinfo {author} {\bibfnamefont {P.}~\bibnamefont
  {Delugas}}, \bibinfo {author} {\bibfnamefont {R.~A.~DiStasio}\ \bibnamefont
  {Jr}}, \bibinfo {author} {\bibfnamefont {A.}~\bibnamefont {Ferretti}},
  \bibinfo {author} {\bibfnamefont {A.}~\bibnamefont {Floris}}, \bibinfo
  {author} {\bibfnamefont {G.}~\bibnamefont {Fratesi}}, \bibinfo {author}
  {\bibfnamefont {G.}~\bibnamefont {Fugallo}}, \bibinfo {author} {\bibfnamefont
  {R.}~\bibnamefont {Gebauer}}, \bibinfo {author} {\bibfnamefont
  {U.}~\bibnamefont {Gerstmann}}, \bibinfo {author} {\bibfnamefont
  {F.}~\bibnamefont {Giustino}}, \bibinfo {author} {\bibfnamefont
  {T.}~\bibnamefont {Gorni}}, \bibinfo {author} {\bibfnamefont {J}~\bibnamefont
  {Jia}}, \bibinfo {author} {\bibfnamefont {M.}~\bibnamefont {Kawamura}},
  \bibinfo {author} {\bibfnamefont {H.-Y.}\ \bibnamefont {Ko}}, \bibinfo
  {author} {\bibfnamefont {A.}~\bibnamefont {Kokalj}}, \bibinfo {author}
  {\bibfnamefont {E.}~\bibnamefont {K\"{u}c\"{u}kbenli}}, \bibinfo {author}
  {\bibfnamefont {M.}~\bibnamefont {Lazzeri}}, \bibinfo {author} {\bibfnamefont
  {M.}~\bibnamefont {Marsili}}, \bibinfo {author} {\bibfnamefont
  {N.}~\bibnamefont {Marzari}}, \bibinfo {author} {\bibfnamefont
  {F.}~\bibnamefont {Mauri}}, \bibinfo {author} {\bibfnamefont {N.~L.}\
  \bibnamefont {Nguyen}}, \bibinfo {author} {\bibfnamefont {H.-V.}\
  \bibnamefont {Nguyen}}, \bibinfo {author} {\bibfnamefont {A.~Otero}\
  \bibnamefont {de-la Roza}}, \bibinfo {author} {\bibfnamefont
  {L.}~\bibnamefont {Paulatto}}, \bibinfo {author} {\bibfnamefont
  {S.}~\bibnamefont {Ponc\'{e}}}, \bibinfo {author} {\bibfnamefont
  {D.}~\bibnamefont {Rocca}}, \bibinfo {author} {\bibfnamefont
  {R.}~\bibnamefont {Sabatini}}, \bibinfo {author} {\bibfnamefont
  {B.}~\bibnamefont {Santra}}, \bibinfo {author} {\bibfnamefont
  {M.}~\bibnamefont {Schlipf}}, \bibinfo {author} {\bibfnamefont {A.~P.}\
  \bibnamefont {Seitsonen}}, \bibinfo {author} {\bibfnamefont {A.}~\bibnamefont
  {Smogunov}}, \bibinfo {author} {\bibfnamefont {I.}~\bibnamefont {Timrov}},
  \bibinfo {author} {\bibfnamefont {T.}~\bibnamefont {Thonhauser}}, \bibinfo
  {author} {\bibfnamefont {P.}~\bibnamefont {Umari}}, \bibinfo {author}
  {\bibfnamefont {N.}~\bibnamefont {Vast}}, \bibinfo {author} {\bibfnamefont
  {X.}~\bibnamefont {Wu}}, \ and\ \bibinfo {author} {\bibfnamefont
  {S.}~\bibnamefont {Baroni}},\ }\href@noop {} {\bibfield  {journal} {\bibinfo
  {journal} {J. Phys.: Condens. Matter}\ }\textbf {\bibinfo {volume} {21}},\
  \bibinfo {pages} {395502} (\bibinfo {year} {2009})}\BibitemShut {NoStop}%
\bibitem [{\citenamefont {Giannozzi}\ \emph {et~al.}(2017)\citenamefont
  {Giannozzi}, \citenamefont {Baroni}, \citenamefont {Bonini}, \citenamefont
  {Calandra}, \citenamefont {Car}, \citenamefont {Cavazzoni}, \citenamefont
  {Ceresoli}, \citenamefont {Chiarotti}, \citenamefont {Cococcioni},
  \citenamefont {amd A.~Dal~Corso}, \citenamefont {Fabris}, \citenamefont
  {Fratesi}, \citenamefont {de~Gironcoli}, \citenamefont {Gebauer},
  \citenamefont {Gerstmann}, \citenamefont {Gougoussis}, \citenamefont
  {Kokalj}, \citenamefont {Lazzeri}, \citenamefont {Martin-Samos},
  \citenamefont {Marzari}, \citenamefont {Mauri}, \citenamefont {Mazzarello},
  \citenamefont {Paolini}, \citenamefont {Pasquarello}, \citenamefont
  {Paulatto}, \citenamefont {Sbraccia}, \citenamefont {Scandolo}, \citenamefont
  {Sclauzero}, \citenamefont {Seitsonen}, \citenamefont {Smogunov},
  \citenamefont {Umari},\ and\ \citenamefont
  {Wentzcovitch}}]{JPhys_CM_29_465901_2017-2}%
  \BibitemOpen
  \bibfield  {author} {\bibinfo {author} {\bibfnamefont {P.}~\bibnamefont
  {Giannozzi}}, \bibinfo {author} {\bibfnamefont {S.}~\bibnamefont {Baroni}},
  \bibinfo {author} {\bibfnamefont {N.}~\bibnamefont {Bonini}}, \bibinfo
  {author} {\bibfnamefont {M.}~\bibnamefont {Calandra}}, \bibinfo {author}
  {\bibfnamefont {R.}~\bibnamefont {Car}}, \bibinfo {author} {\bibfnamefont
  {C.}~\bibnamefont {Cavazzoni}}, \bibinfo {author} {\bibfnamefont
  {D.}~\bibnamefont {Ceresoli}}, \bibinfo {author} {\bibfnamefont {G.~L.}\
  \bibnamefont {Chiarotti}}, \bibinfo {author} {\bibfnamefont {M.}~\bibnamefont
  {Cococcioni}}, \bibinfo {author} {\bibfnamefont {I.~Dabo}\ \bibnamefont {amd
  A.~Dal~Corso}}, \bibinfo {author} {\bibfnamefont {S.}~\bibnamefont {Fabris}},
  \bibinfo {author} {\bibfnamefont {G.}~\bibnamefont {Fratesi}}, \bibinfo
  {author} {\bibfnamefont {S.}~\bibnamefont {de~Gironcoli}}, \bibinfo {author}
  {\bibfnamefont {R.}~\bibnamefont {Gebauer}}, \bibinfo {author} {\bibfnamefont
  {U.}~\bibnamefont {Gerstmann}}, \bibinfo {author} {\bibfnamefont
  {C.}~\bibnamefont {Gougoussis}}, \bibinfo {author} {\bibfnamefont
  {A.}~\bibnamefont {Kokalj}}, \bibinfo {author} {\bibfnamefont
  {M.}~\bibnamefont {Lazzeri}}, \bibinfo {author} {\bibfnamefont
  {L.}~\bibnamefont {Martin-Samos}}, \bibinfo {author} {\bibfnamefont
  {N.}~\bibnamefont {Marzari}}, \bibinfo {author} {\bibfnamefont
  {F.}~\bibnamefont {Mauri}}, \bibinfo {author} {\bibfnamefont
  {R.}~\bibnamefont {Mazzarello}}, \bibinfo {author} {\bibfnamefont
  {S.}~\bibnamefont {Paolini}}, \bibinfo {author} {\bibfnamefont
  {A.}~\bibnamefont {Pasquarello}}, \bibinfo {author} {\bibfnamefont
  {L.}~\bibnamefont {Paulatto}}, \bibinfo {author} {\bibfnamefont
  {C.}~\bibnamefont {Sbraccia}}, \bibinfo {author} {\bibfnamefont
  {S.}~\bibnamefont {Scandolo}}, \bibinfo {author} {\bibfnamefont
  {G.}~\bibnamefont {Sclauzero}}, \bibinfo {author} {\bibfnamefont {A.~P.}\
  \bibnamefont {Seitsonen}}, \bibinfo {author} {\bibfnamefont {A.}~\bibnamefont
  {Smogunov}}, \bibinfo {author} {\bibfnamefont {P.}~\bibnamefont {Umari}}, \
  and\ \bibinfo {author} {\bibfnamefont {R.~M.}\ \bibnamefont {Wentzcovitch}},\
  }\href@noop {} {\bibfield  {journal} {\bibinfo  {journal} {J. Phys.: Condens.
  Matter}\ }\textbf {\bibinfo {volume} {29}},\ \bibinfo {pages} {465901}
  (\bibinfo {year} {2017})}\BibitemShut {NoStop}%
\bibitem [{\citenamefont {Perdew}\ \emph {et~al.}(1996)\citenamefont {Perdew},
  \citenamefont {Burke},\ and\ \citenamefont
  {Ernzerhof}}]{PhysRevLett_77_3865_1996-2}%
  \BibitemOpen
  \bibfield  {author} {\bibinfo {author} {\bibfnamefont {J.~P.}\ \bibnamefont
  {Perdew}}, \bibinfo {author} {\bibfnamefont {K.}~\bibnamefont {Burke}}, \
  and\ \bibinfo {author} {\bibfnamefont {M.}~\bibnamefont {Ernzerhof}},\
  }\bibfield  {title} {\enquote {\bibinfo {title} {Generalized gradient
  approximation made simple},}\ }\href@noop {} {\bibfield  {journal} {\bibinfo
  {journal} {Phys. Rev. Lett.}\ }\textbf {\bibinfo {volume} {77}},\ \bibinfo
  {pages} {3865} (\bibinfo {year} {1996})}\BibitemShut {NoStop}%
\bibitem [{\citenamefont {Hamann}\ \emph {et~al.}(1979)\citenamefont {Hamann},
  \citenamefont {Schl\"{u}ter},\ and\ \citenamefont
  {Chiang}}]{PhysRevLett_43_1494-1497_1979-2}%
  \BibitemOpen
  \bibfield  {author} {\bibinfo {author} {\bibfnamefont {D.}~\bibnamefont
  {Hamann}}, \bibinfo {author} {\bibfnamefont {M.}~\bibnamefont
  {Schl\"{u}ter}}, \ and\ \bibinfo {author} {\bibfnamefont {C.}~\bibnamefont
  {Chiang}},\ }\bibfield  {title} {\enquote {\bibinfo {title} {Norm-conserving
  pseudopotentials},}\ }\href@noop {} {\bibfield  {journal} {\bibinfo
  {journal} {Phys. Rev. Lett.}\ }\textbf {\bibinfo {volume} {43}},\ \bibinfo
  {pages} {1494--1497} (\bibinfo {year} {1979})}\BibitemShut {NoStop}%
\bibitem [{\citenamefont {Bachelet}\ \emph {et~al.}(1982)\citenamefont
  {Bachelet}, \citenamefont {Hamann},\ and\ \citenamefont
  {Schl\"{u}ter}}]{PhysRevB_26_4199-4228_1982-2}%
  \BibitemOpen
  \bibfield  {author} {\bibinfo {author} {\bibfnamefont {G.}~\bibnamefont
  {Bachelet}}, \bibinfo {author} {\bibfnamefont {D.}~\bibnamefont {Hamann}}, \
  and\ \bibinfo {author} {\bibfnamefont {M.}~\bibnamefont {Schl\"{u}ter}},\
  }\bibfield  {title} {\enquote {\bibinfo {title} {Pseudopotentials that work:
  From h to pu},}\ }\href@noop {} {\bibfield  {journal} {\bibinfo  {journal}
  {Phys. Rev. B}\ }\textbf {\bibinfo {volume} {26}},\ \bibinfo {pages}
  {4199--4228} (\bibinfo {year} {1982})}\BibitemShut {NoStop}%
\bibitem [{\citenamefont {Grimme}(2006)}]{J_Comp_Chem_27_1787_2006-2}%
  \BibitemOpen
  \bibfield  {author} {\bibinfo {author} {\bibfnamefont {S.}~\bibnamefont
  {Grimme}},\ }\bibfield  {title} {\enquote {\bibinfo {title} {Semiempirical
  gga‐type density functional constructed with a long‐range dispersion
  correction},}\ }\href@noop {} {\bibfield  {journal} {\bibinfo  {journal} {J.
  Comp. Chem.}\ }\textbf {\bibinfo {volume} {27}},\ \bibinfo {pages} {1787}
  (\bibinfo {year} {2006})}\BibitemShut {NoStop}%
\bibitem [{\citenamefont {Deslippe}\ \emph {et~al.}(2012)\citenamefont
  {Deslippe}, \citenamefont {Samsonidze}, \citenamefont {Strubbe},
  \citenamefont {Jain}, \citenamefont {Cohen},\ and\ \citenamefont
  {Louie}}]{Com_Phys_Comm_183_1269_2012-2}%
  \BibitemOpen
  \bibfield  {author} {\bibinfo {author} {\bibfnamefont {J.}~\bibnamefont
  {Deslippe}}, \bibinfo {author} {\bibfnamefont {G.}~\bibnamefont
  {Samsonidze}}, \bibinfo {author} {\bibfnamefont {D.~A.}\ \bibnamefont
  {Strubbe}}, \bibinfo {author} {\bibfnamefont {M.}~\bibnamefont {Jain}},
  \bibinfo {author} {\bibfnamefont {M.~L.}\ \bibnamefont {Cohen}}, \ and\
  \bibinfo {author} {\bibfnamefont {S.~G.}\ \bibnamefont {Louie}},\ }\bibfield
  {title} {\enquote {\bibinfo {title} {Berkeleygw: A massively parallel
  computer package for the calculation of the quasiparticle and optical
  properties of materials and nanostructures},}\ }\href@noop {} {\bibfield
  {journal} {\bibinfo  {journal} {Comput. Phys. Commun.}\ }\textbf {\bibinfo
  {volume} {183}},\ \bibinfo {pages} {1269} (\bibinfo {year}
  {2012})}\BibitemShut {NoStop}%
\bibitem [{\citenamefont {Hybertsen}\ and\ \citenamefont
  {Louie}(1986)}]{PhysRevB_34_5390_1986-2}%
  \BibitemOpen
  \bibfield  {author} {\bibinfo {author} {\bibfnamefont {M.}~\bibnamefont
  {Hybertsen}}\ and\ \bibinfo {author} {\bibfnamefont {S.}~\bibnamefont
  {Louie}},\ }\bibfield  {title} {\enquote {\bibinfo {title} {Electron
  correlation in semiconductors and insulators: Band gaps and quasiparticle
  energies},}\ }\href@noop {} {\bibfield  {journal} {\bibinfo  {journal} {Phys.
  Rev. B}\ }\textbf {\bibinfo {volume} {34}},\ \bibinfo {pages} {5390--5413}
  (\bibinfo {year} {1986})}\BibitemShut {NoStop}%
\bibitem [{\citenamefont {Deslippe}\ \emph {et~al.}(2013)\citenamefont
  {Deslippe}, \citenamefont {Samsonidze}, \citenamefont {Jain}, \citenamefont
  {Cohen},\ and\ \citenamefont {Louie}}]{PhysRevB_87_165124_2013-2}%
  \BibitemOpen
  \bibfield  {author} {\bibinfo {author} {\bibfnamefont {J.}~\bibnamefont
  {Deslippe}}, \bibinfo {author} {\bibfnamefont {G.}~\bibnamefont
  {Samsonidze}}, \bibinfo {author} {\bibfnamefont {M.}~\bibnamefont {Jain}},
  \bibinfo {author} {\bibfnamefont {M.~L.}\ \bibnamefont {Cohen}}, \ and\
  \bibinfo {author} {\bibfnamefont {S.~G.}\ \bibnamefont {Louie}},\ }\bibfield
  {title} {\enquote {\bibinfo {title} {Coulomb-hole summations and energies for
  $gw$ calculations with limited number of empty orbitals: A modified static
  remainder approach},}\ }\href@noop {} {\bibfield  {journal} {\bibinfo
  {journal} {Phys. Rev. B}\ }\textbf {\bibinfo {volume} {87}},\ \bibinfo
  {pages} {165124} (\bibinfo {year} {2013})}\BibitemShut {NoStop}%
\bibitem [{\citenamefont {Ismail-Beigi}(2006)}]{PhysRevB_73_233103_2006-2}%
  \BibitemOpen
  \bibfield  {author} {\bibinfo {author} {\bibfnamefont {S.}~\bibnamefont
  {Ismail-Beigi}},\ }\bibfield  {title} {\enquote {\bibinfo {title} {Truncation
  of periodic image interactions for confined systems},}\ }\href@noop {}
  {\bibfield  {journal} {\bibinfo  {journal} {Phys. Rev. B}\ }\textbf {\bibinfo
  {volume} {73}},\ \bibinfo {pages} {233103} (\bibinfo {year}
  {2006})}\BibitemShut {NoStop}%
\bibitem [{\citenamefont {Qiu}\ \emph {et~al.}(2013)\citenamefont {Qiu},
  \citenamefont {da~Jornada},\ and\ \citenamefont
  {Louie}}]{PhysRevLett_111_216805_2013-2}%
  \BibitemOpen
  \bibfield  {author} {\bibinfo {author} {\bibfnamefont {D.~Y.}\ \bibnamefont
  {Qiu}}, \bibinfo {author} {\bibfnamefont {F.~H.}\ \bibnamefont {da~Jornada}},
  \ and\ \bibinfo {author} {\bibfnamefont {S.~G.}\ \bibnamefont {Louie}},\
  }\bibfield  {title} {\enquote {\bibinfo {title} {Optical spectrum of
  $\textrm{MoS}_{2}$: Many-body effects and diversity of excition states},}\
  }\href@noop {} {\bibfield  {journal} {\bibinfo  {journal} {Phys. Rev. Lett.}\
  }\textbf {\bibinfo {volume} {111}},\ \bibinfo {pages} {216805} (\bibinfo
  {year} {2013})}\BibitemShut {NoStop}%
\bibitem [{\citenamefont {Yu}\ and\ \citenamefont
  {Zunger}(2012)}]{PhysRevLett_108_068701_2012-2}%
  \BibitemOpen
  \bibfield  {author} {\bibinfo {author} {\bibfnamefont {L.}~\bibnamefont
  {Yu}}\ and\ \bibinfo {author} {\bibfnamefont {A.}~\bibnamefont {Zunger}},\
  }\bibfield  {title} {\enquote {\bibinfo {title} {Identification of potential
  photovoltaic absorbers based on first-principles spectroscopic screening of
  materials},}\ }\href@noop {} {\bibfield  {journal} {\bibinfo  {journal}
  {Phys. Rev. Lett.}\ }\textbf {\bibinfo {volume} {108}},\ \bibinfo {pages}
  {068701} (\bibinfo {year} {2012})}\BibitemShut {NoStop}%
\bibitem [{\citenamefont {Choudhary}\ \emph {et~al.}(2019)\citenamefont
  {Choudhary}, \citenamefont {Bercx}, \citenamefont {Jiang}, \citenamefont
  {Pachter}, \citenamefont {Lamoen},\ and\ \citenamefont
  {Tavazza}}]{acs.chemmater.9b02166-2}%
  \BibitemOpen
  \bibfield  {author} {\bibinfo {author} {\bibfnamefont {Kamal}\ \bibnamefont
  {Choudhary}}, \bibinfo {author} {\bibfnamefont {Marnik}\ \bibnamefont
  {Bercx}}, \bibinfo {author} {\bibfnamefont {Jie}\ \bibnamefont {Jiang}},
  \bibinfo {author} {\bibfnamefont {Ruth}\ \bibnamefont {Pachter}}, \bibinfo
  {author} {\bibfnamefont {Dirk}\ \bibnamefont {Lamoen}}, \ and\ \bibinfo
  {author} {\bibfnamefont {Francesca}\ \bibnamefont {Tavazza}},\ }\bibfield
  {title} {\enquote {\bibinfo {title} {Accelerated discovery of efficient solar
  cell materials using quantum and machine-learning methods},}\ }\href@noop {}
  {\bibfield  {journal} {\bibinfo  {journal} {Chemistry of Materials}\ }\textbf
  {\bibinfo {volume} {31}},\ \bibinfo {pages} {5900--5908} (\bibinfo {year}
  {2019})}\BibitemShut {NoStop}%
\bibitem [{AM1()}]{AM1.5-2}%
  \BibitemOpen
  \href@noop {} {\enquote {\bibinfo {title} {Reference solar spectral
  irradiance: Air mass 1.5},}\ }\bibinfo {howpublished}
  {\url{https://rredc.nrel.gov/solar//spectra/am1.5/}}\BibitemShut {NoStop}%
\bibitem [{\citenamefont {Chae}\ and\ \citenamefont
  {Son}(2019)}]{NanoLett_19_2694_2019-2}%
  \BibitemOpen
  \bibfield  {author} {\bibinfo {author} {\bibfnamefont {Kisung}\ \bibnamefont
  {Chae}}\ and\ \bibinfo {author} {\bibfnamefont {Young-Woo}\ \bibnamefont
  {Son}},\ }\bibfield  {title} {\enquote {\bibinfo {title} {A new family of
  two-dimensional crystals: Open-framework $\textrm{T}_{3}\textrm{X}$
  ($\textrm{T = C, Si, Ge, Sn}$; $\textrm{X = O, S, Se, Te}$) compounds with
  tetrahedral bonding},}\ }\href@noop {} {\bibfield  {journal} {\bibinfo
  {journal} {Nano. Lett.}\ }\textbf {\bibinfo {volume} {19}},\ \bibinfo {pages}
  {2694--2699} (\bibinfo {year} {2019})}\BibitemShut {NoStop}%
\bibitem [{\citenamefont {Vurgaftman}\ \emph {et~al.}(2001)\citenamefont
  {Vurgaftman}, \citenamefont {Meyer},\ and\ \citenamefont
  {Ram-Mohan}}]{doi:10.1063/1.1368156-2}%
  \BibitemOpen
  \bibfield  {author} {\bibinfo {author} {\bibfnamefont {I.}~\bibnamefont
  {Vurgaftman}}, \bibinfo {author} {\bibfnamefont {J.~R.}\ \bibnamefont
  {Meyer}}, \ and\ \bibinfo {author} {\bibfnamefont {L.~R.}\ \bibnamefont
  {Ram-Mohan}},\ }\bibfield  {title} {\enquote {\bibinfo {title} {Band
  parameters for iii–v compound semiconductors and their alloys},}\
  }\href@noop {} {\bibfield  {journal} {\bibinfo  {journal} {Journal of Applied
  Physics}\ }\textbf {\bibinfo {volume} {89}},\ \bibinfo {pages} {5815--5875}
  (\bibinfo {year} {2001})}\BibitemShut {NoStop}%
\end{thebibliography}

\end{document}